\documentclass[12pt]{article}

\usepackage{amssymb,amsthm,amscd, amsbsy, array}

\let\ssection=\section
\renewcommand{\section}{\setcounter{equation}{0}\ssection}

\newcommand{\Ad}{\mathrm{Ad}}

\newcommand{\rA}{\mathrm{A}}

\newcommand{\bone}{\boldsymbol{1}}

\newcommand{\bc}{{\bf c}}

\newcommand{\Coad}{\mathrm{Coad}}

\newcommand{\rdiv}{\mathrm{div}}

\newcommand{\rE}{{\mathrm{E}}}

\newcommand{\Ein}{{\mathrm{Ein}}}

\newcommand{\bg}{{\bf g}}

\newcommand{\rg}{\mathrm{g}}

\newcommand{\Gal}{\mathrm{Gal}}
\newcommand{\fg}{\mathfrak{g}}

\newcommand{\bgamma}{\boldsymbol{\gamma}}

\newcommand{\GL}{{\mathrm{GL}}}
\newcommand{\grad}{{\mathbf{grad}}}

\newcommand{\fh}{\mathfrak{h}}

\newcommand{\bell}{\boldsymbol{\ell}}
\newcommand{\lambdabar}{{\lambda\!\!\!^{-}}}

\newcommand{\cM}{\mathcal{M}}

\newcommand{\bn}{{\bf n}}

\newcommand{\rO}{\mathrm{O}}
\newcommand{\cO}{\mathcal{O}}
\newcommand{\bomega}{\boldsymbol{\omega}}

\newcommand{\bp}{{\bf p}}
\newcommand{\wbp}{\widehat{{\bf p}}}
\newcommand{\bP}{{\bf P}}

\newcommand{\bq}{{\bf q}}

\newcommand{\bbR}{\mathbb{R}}

\newcommand{\Ric}{\mathrm{Ric}}

\newcommand{\sign}{\mathrm{sign}}
\newcommand{\SE}{\mathrm{SE}}

\newcommand{\SO}{\mathrm{SO}}
\newcommand{\se}{\mathfrak{se}}

\newcommand{\Surf}{\mathrm{Surf}}

\newcommand{\Tr}{\mathrm{Tr}}

\newcommand{\bu}{{\bf u}}

\newcommand{\bv}{{\bf v}}

\newcommand{\vol}{\mathrm{vol}}

\newcommand{\bw}{{\bf w}}

\newcommand{\bx}{{\bf x}}

\newcommand{\bz}{{\bf z}}

\newcommand{\bbZ}{\mathbb{Z}}

\newcommand{\const}{\mathop{\rm const.}\nolimits}
\newcommand{\half}{\frac{1}{2}}

\newcommand{\la}{{\langle}}
\newcommand{\ra}{{\rangle}}

\begin{document}

\baselineskip=22pt

\oddsidemargin .1truein
\newtheorem{thm}{Theorem}[section]
\newtheorem{lem}[thm]{Lemma}
\newtheorem{cor}[thm]{Corollary}
\newtheorem{pro}[thm]{Proposition}
\newtheorem{ex}[thm]{Example}
\newtheorem{rmk}[thm]{Remark}
\newtheorem{defi}[thm]{Definition}

%%%%%%%%%%%%%%%%%%%%%%%%%%%%%%%%%%%%%%%%%%%%%%%%%%%%%%%%%%%%%%%%%%%%%%%%%%%%%%%%%%%%
%%%%%%%%%%%%%%%%%%%%%%%%%%%%%%%%%%%%%%%%%%%%%%%%%%%%%%%%%%%%%%%%%%%%%%%%%%%%%%%%%%%%
\title{Geometrical Spinoptics\\
and\\
the Optical Hall Effect}
%%%%%%%%%%%%%%%%%%%%%%%%%%%%%%%%%%%%%%%%%%%%%%%%%%%%%%%%%%%%%%%%%%%%%%%%%%%%%%%%%%%%
%%%%%%%%%%%%%%%%%%%%%%%%%%%%%%%%%%%%%%%%%%%%%%%%%%%%%%%%%%%%%%%%%%%%%%%%%%%%%%%%%%%%

\author{
C. DUVAL\footnote{mailto: duval@cpt.univ-mrs.fr}\\
Centre de Physique Th\'eorique, CNRS, 
Luminy, Case 907\\ 
F-13288 Marseille Cedex 9 (France)\footnote{ 
UMR 6207 du CNRS associ\'ee aux 
Universit\'es d'Aix-Marseille I et II et Universit\'e du Sud Toulon-Var; Laboratoire 
affili\'e \`a la FRUMAM-FR2291
}
\and
Z.~HORV\'ATH\footnote{mailto: zalanh@ludens.elte.hu}\\
Institute for Theoretical Physics, E\"otv\"os University\\
P\'azm\'any P. s\'et\'any 1/A\\
H-1117 BUDAPEST (Hungary)
\and
%\\
P.~A.~HORV\'ATHY\footnote{mailto: horvathy@lmpt.univ-tours.fr}\\
Laboratoire de Math\'ematiques et de Physique Th\'eorique\\
Universit\'e de Tours, Parc de Grandmont\\
F-37 200 TOURS (France)
}

%\date{12 September 2005}
\date{}

\maketitle

\thispagestyle{empty}

\begin{abstract}
Geometrical optics is extended so as to provide a model for spinning light rays via the coadjoint orbits of the Euclidean group characterized by color and spin. This leads to a theory of ``geometrical spinoptics'' in refractive media. Symplectic scattering yields generalized Snell-Descartes laws that include the recently discovered optical Hall effect.
\end{abstract}

%\bigskip
%\noindent
%\textbf{Keywords:} Geometrical optics, polarization, coadjoint orbits, Hall effect.

%\vskip1cm
\bigskip


%Keywords: Geometrical optics, Euclidean coadjoint orbits, Minimal coupling, Symplectic scattering.

Preprint : CPT-2005/P.052
%:
\hfill\texttt{arXiv:math-ph/0509031}
\newpage

%\tableofcontents\newpage

%%%%%%%%%%%%%%%%%%%%%%%%%%%%%%%%%%%%%%%%%%%%%%%%%%%%%%%%%%%%%%%%%%%%%%%%%%%%%%%%%%%%
%%%%%%%%%%%%%%%%%%%%%%%%%%%%%%%%%%%%%%%%%%%%%%%%%%%%%%%%%%%%%%%%%%%%%%%%%%%%%%%%%%%%
\section{Introduction}\label{Introduction}
%%%%%%%%%%%%%%%%%%%%%%%%%%%%%%%%%%%%%%%%%%%%%%%%%%%%%%%%%%%%%%%%%%%%%%%%%%%%%%%%%%%%
%%%%%%%%%%%%%%%%%%%%%%%%%%%%%%%%%%%%%%%%%%%%%%%%%%%%%%%%%%%%%%%%%%%%%%%%%%%%%%%%%%%%

The mathematical foundation of optics certainly goes back to Euclid's Optics with the observation that light travels along straight lines, namely along the geodesics of Euclidean space. Geometrical optics has, since the breakthrough provided by the Fermat principle of least optical path, proved extremely useful as a theory of light, prior to Maxwellian wave optics. Suffice it to mention its relevance in the design of optical (as well as electronic!) devices using lenses, mirrors, etc., its importance in the understanding of caustics~\cite{Arn}, in the modelling of optical aberrations, etc. Geometrical optics has, since then, been recognized as a ``semi-classical'' limit of wave optics with small parameter $\lambdabar$ (where~$\lambda$ is a typical wavelength); it has nevertheless been constantly considered as a self-consistent theory for light rays, borrowing much from differential geometry, and, more specifically, from Riemannian and symplectic geometries. Geometrical optics provides, indeed, a beautiful link between both previously mentioned geometries: (i) light travels along (oriented) geodesics of an optical medium, a $3$-dimensional manifold whose Riemannian structure is defined by the refractive index, (ii) the set of all such geodesics is naturally endowed with a structure of $4$-dimensional symplectic manifold. It is this  duality that will serve as an Ariadne's thread in our subsequent extension of geometrical optics.

In addition, it is nowadays a well-established experimental fact that trajectories of light beams in inhomogeneous media slightly depart from those enacted by the Fermat principle. This class of effects predicted by Fedorov, fifty years ago, has since then received numerous theoretical interpretations that go back to work of Costa de Beauregard~\cite{Cos}, Boulware \cite{Bou}, among many a fundamental and more recent contribution (see, e.g.,~\cite{BB04,BB05} for a brief historical account with an updated list of references regarding both theory and experiment). One the measured effects is the ``Magnus Effect for light'' that describes how trajectories depend upon the polarization state of the beam in weakly inhomogeneous media. Another related effect is the so-called ``Optical Hall Effect'' (OHE) which has lately received special attention and is associated with a transverse shift of the position of a photonic wave packet at the interface separating two media of different refractive indices. (The OHE actually bears strong resemblance with the Hall effect governing, in two-dimensional conducting samples, the electronic transport, transverse to the electric field and an applied external magnetic field.) Such a shift, transverse to the incidence plane, has already been experimentally measured by Imbert \cite{Imb} in the case of total reflection. As to the transverse shift for partial reflection and refraction, it is currently under highly active investigation from both a theoretical and an experimental perspective; see, e.g., \cite{BB04,BB05} and \cite{OMN}. Let us also mention the recently discovered phenomenon of magneto-transverse light diffusion in Faraday-active dielectric media~\cite{Tig}, the ``Photonic Hall Effect'' (PHE), see \cite{WSRLvT}, which is clearly a spin-induced effect.

The need for a generalization of geometrical optics which would consistently include polarization effects hence became mandatory. Various approaches, including a full-fledged computation of refraction and reflection of arbitrarily polarized Gaussian electromagnetic wave packets \cite{BB05}, have been put forward. Of particular interest are recent extensions of geometrical optics, within a Maxwellian context, using a certain ``Berry connection'' whose curvature (in momentum space) yields a modification of the Fermat equations of motion for polarized light beams~\cite{LZ,BB04,BB04bis,BF, OMN}; see also~\cite{BM,Bli,GBM}. A quasi-classical formula for the above-mentioned transverse shift of polarized light beams has also been proposed by Onoda, Murakami, and Nagaosa~\cite{OMN}, together with an experimental set up using photonic crystals in order to reveal the OHE for reflected and refracted light beams. Nevertheless, no consensus seems to have emerged so far regarding a clear-cut theory of geometrical optics including polarization.

\goodbreak

Our standpoint is to take advantage of the fact that geometrical optics is fundamentally related to Euclidean geometry since oriented straight lines (light rays) actually consist of specific coadjoint orbits of the Euclidean group, $\rE(3)$. An extended theory of geometrical optics should therefore be expected to emerge from the same Euclidean geometry, more precisely from the consideration of the other $\rE(3)$-coadjoint orbits which seem to have been overlooked by physicists. The purpose of this article is therefore to view and exploit the generic $\rE(3)$-coadjoint orbits, carrying color and spin according to Souriau's classification \cite{Sou}, as the sets of ``free'' colored and circularly polarized light rays. We then introduce a natural adaptation of the Fermat prescription to the spin case, inspired by the prescription of ``minimal coupling'' to a curved metric used in general relativity, see, e.g., \cite{DFS,Kun,Sou,Ste}. This leads us in a straightforward fashion to a theory describing the trajectories of spinning light rays in arbitrary dielectric media. We name this theory ``geometrical spinoptics''. 

The present article is organized as follows.

Section \ref{SecEuclideanOptics} is devoted to a general overview of the geometry of the set of light rays with color and spin as coadjoint orbits of the group $\SE(3)$ of Euclidean orientation-preserving\footnote{Considering, here, the neutral component of the Euclidean group is a technicality which will be discussed and justified below.} isometries of Euclidean $3$-dimensional space space. It should be highlighted that the generic coadjoint orbits, with topology $TS^2$, are automatically endowed with a ``twisted'' cotangent symplectic structure which describes spin---or polarization. There is, hence, no need to introduce a ``Berry curvature'' which is, in a sense, already encoded in the above-mentioned geometrical twist. The relationship of these Euclidean coadjoint orbits to the massless, spinning, coadjoint orbits of the Poincar\'e and Galilei groups is spelled out explicitly.

In Section \ref{SecSpinOptics}, we generalize the Fermat prescription so as to describe the pre\-symplectic structure of colored and spinning light rays in an arbitrary refractive medium characterized by a dielectric tensor defining a Riemannian metric. This prescription happens to be akin to the ``minimal coupling'' procedure used, in general relativity, to account for the geodesic deviation of spinning particles due to tidal forces generated by the interaction of spin and spacetime curvature. The general equations for spinoptics (\ref{kersigmaSTMg}) are then derived. The special case of an isotropic and inhomogeneous medium, of refractive index $n$, is worked out in great detail; we recover, upon linearization of the foliation (\ref{kersigmaSpinFermatBis}) around the value $n=\const$, a system of equations of motion for polarized light rays proposed in \cite{OMN}.

\goodbreak

Section \ref{SecSD-OHE} gives us the opportunity to derive, in the present framework, the Snell-Descartes laws (\ref{SnellDescartes}) for spinoptics, dictated by Souriau's symplectic scattering. We explicitly compute the form of the reflection and refraction (local) symplecto\-morphism between ``in'' and ``out'' polarized optical states. A novel and subtle phenomenon naturally stems from our geometrical treatment, viz., the fact that scattered spinning light rays are actually shifted, transversally to the plane of incidence. The formula (\ref{ShiftBis}) we derive for this transverse shift agrees with the one proposed in~\cite{OMN} to describe the OHE from a different standpoint. This shift is then analyzed for left-handed media, providing a plausible mechanical interpretation of the ``perfectness'' of superlenses \cite{Pen}.

At last, in Section \ref{Conclusion}, we sum up the content of this article, and conclude by presenting several generalizations in prospect, e.g., a theory of geometrical spinoptics for Faraday-active media.

A companion article \cite{DHH} provides an overview of geometrical spinoptics and discusses its relation to other physically oriented approaches found in the recent literature.

\medskip
\noindent
\textbf{Acknowledgment}: We are indebted to K. Y. Bliokh for fruitful correspondence.

%%%%%%%%%%%%%%%%%%%%%%%%%%%%%%%%%%%%%%%%%%%%%%%%%%%%%%%%%%%%%%%%%%%%%%%%%%%%%%%%%%%%
%%%%%%%%%%%%%%%%%%%%%%%%%%%%%%%%%%%%%%%%%%%%%%%%%%%%%%%%%%%%%%%%%%%%%%%%%%%%%%%%%%%%
\section{Geometrical optics and the Euclidean group}\label{SecEuclideanOptics}
%%%%%%%%%%%%%%%%%%%%%%%%%%%%%%%%%%%%%%%%%%%%%%%%%%%%%%%%%%%%%%%%%%%%%%%%%%%%%%%%%%%%
%%%%%%%%%%%%%%%%%%%%%%%%%%%%%%%%%%%%%%%%%%%%%%%%%%%%%%%%%%%%%%%%%%%%%%%%%%%%%%%%%%%%

An oriented straight line, $\xi$, in Euclidean (affine) space $(E^3, \la\,\cdot\,,\,\cdot\,\ra)$ is determined by its direction, i.e., a vector $\bu\in\bbR^3$ of unit length and an arbitrary point $M\in\xi$. Should an origin, $O\in{}E^3$, be chosen, we may take the vector $\bq=M-O$ to be orthogonal to~$\bu$. The set of oriented, non parametrized, straight lines is plainly the smooth manifold
\begin{equation}
\cM=\{\xi=(\bq,\bu)\in\bbR^3\times\bbR^3\,\vert\,\la\bu,\bu\ra=1,\la\bu,\bq\ra=0\},
\label{StraightLines}
\end{equation}
i.e., the tangent bundle $\cM\cong{}TS^2$ of the round sphere $S^2$.

\goodbreak

Now $TS^2$ has been recognized \cite{Sou} as a coadjoint orbit of the group, $\rE(3)$, of Euclidean isometries and inherits, as such, an $\rE(3)$-invariant symplectic structure; see also \cite{Igl}.

Let us briefly recall the general construction leading, in particular, to the previous symplectic manifold. Start with the group $\SE(3)=\SO(3)\ltimes\bbR^3$ of orientation-preserving Euclidean isometries whose elements we denote $g=(R,\bx)$. Let $\mu=(\bell,\bp)$ be a point in $\se(3)^*$ where $\se(3)=\bbR^3\ltimes\bbR^3$ is the Lie algebra of $\SE(3)$. The coadjoint representation\footnote{The coadjoint representation, $\Coad$, is defined by $\Coad_g\mu\equiv\mu\circ\Ad_{g^{-1}}$ where $\Ad$ is the adjoint representation.} of $\SE(3)$ reads then $\Coad_{g}(\bell,\bp)=(R\bell+\bx\times{}R\bp,R\bp)$ where~$\times$ stands here for standard cross-product. Obviously $C=\Vert\bp\Vert^2=\la\bp,\bp\ra$ and $C'=\la\bell,\bp\ra$ are invariants of the coadjoint representation. If $C=0$, then $C''=\Vert\bell\Vert^2$ is an extra invariant. These are, in fact, the only invariants and fixing $(C,C')$ or $(C=0,C'')$ yields a single coadjoint orbit \cite{GS,MR,Sou}.

\goodbreak

Consider now, in full generality, a finite-dimensional Lie group $G$ whose Lie algebra is denoted $\fg$. Fix $\mu_0\in\fg^*$ and posit the following $1$-form 
\begin{equation}
\varpi=\mu_0\cdot\vartheta
\label{canonical1form}
\end{equation}
on $G$ where $\vartheta$ stands for the (left-invariant) Maurer-Cartan $1$-form of the group. It is a general fact that $\sigma=d\varpi$ is a presymplectic $2$-form on $G$ which descends as the canonical Kirillov-Kostant-Souriau \cite{Kir,Kos,Sou} symplectic $2$-form, $\omega$, on the coadjoint orbit $\cO_{\mu_0}=\{\mu=\Coad_g(\mu_0)\,\vert\,g\in{}G\}\cong{}G/G_{\mu_0}$ where $G_{\mu_0}$ is the stabilizer  of~$\mu_0\in\fg^*$.

\goodbreak

%%%%%%%%%%%%%%%%%%%%%%%%%%%%%%%%%%%%%%%%%%%%%%%%%%%%%%%%%%%%%%%%%%%%%%%%%%%%%%%%%%%%
\subsection{Colored light rays}
%%%%%%%%%%%%%%%%%%%%%%%%%%%%%%%%%%%%%%%%%%%%%%%%%%%%%%%%%%%%%%%%%%%%%%%%%%%%%%%%%%%%

Specializing this construction to the case $G=\SE(3)$, with $C=p^2$ and $p>0$ together with $C'=0$, we can choose $\mu_0=(\mathbf{0},\bp_0)$ and $\bp_0=(p,0,0)$. 

The associated $1$-form on the group reads then\footnote{We write (\ref{varpi0}) as a useful shorthand for $\varpi=p\,\delta_{ij}u^idx^j$, where $i,j=1,2,3$; the Einstein summation convention is being understood.}
\begin{equation}
\varpi=p\la\bu,d\bx\ra
\label{varpi0}
\end{equation}
where $R=(\bu,\bv,\bw)$ is viewed as an orthonormal, positively oriented, basis of $\bbR^3$. Its exterior derivative, $\sigma=d\varpi$, retains the form
\begin{equation}
\sigma(\delta{g},\delta'{g})=p\left[\la\delta\bu,\delta'\bx\ra-\la\delta'\bu,\delta\bx\ra\right]
\label{sigma0}
\end{equation}
for all $\delta{g},\delta'g\in{}T_g\,\SE(3)$; this $2$-form clearly descends to the spherical tangent bundle $ST\bbR^3=S^2\times\bbR^3$ of $\bbR^3$ described by the pairs $(\bu,\bx)$. 

\goodbreak

Computing the kernel of the latter $2$-form yields 
\begin{equation}
(\delta\bu,\delta\bx)\in\ker(\sigma)
\Longleftrightarrow
\left\{
\begin{array}{rcll}
\delta\bu&=&0\\
\delta\bx&=&\alpha\bu
\end{array}
\right.
\label{kersigma0}
\end{equation}
with $\alpha\in\bbR$. We recognize in (\ref{kersigma0}) the foliation defining the equations of the geodesics of Euclidean space $(E^3, \la\,\cdot\,,\,\cdot\,\ra)$, or, in the context of geometrical optics, the light rays in vacuum.

The first-integrals $\xi=(\bq,\bu)$ of the foliation (\ref{kersigma0}) where 
\begin{equation}
\bq=\bx-\bu\la\bu,\bx\ra
\label{q}
\end{equation}
parametrize the manifold $\cO_{\mu_0}\cong{}TS^2$ of light rays of color $p$, see (\ref{StraightLines}). (This invariant, $p$, of the coadjoint representation of $\rE(3)$ has been coined ``color'' in~\cite{Sou} as $2\pi\hbar/p$ may be interpreted as the Euclidean ``wavelength'' of the light rays.)
 The symplectic form, $\omega$, of the latter manifold is such that $\sigma=(\SE(3)\to\cO_{\mu_0})^*\omega$, viz.,
\begin{equation}
\sigma(\delta{g},\delta'{g})=\omega(\delta\xi,\delta'\xi)=p\left[\la\delta\bu,\delta'\bq\ra-\la\delta'\bu,\delta\bq\ra\right]
\label{omega0}
\end{equation}
or, equivalently, 
\begin{equation}
\omega=d\theta
\qquad
\&%\mathrm{with}
\qquad
\theta=-p\la\bq,d\bu\ra.
\label{alpha0}
\end{equation}
Let us note that the $\SE(3)$-coadjoint orbit $\cO_{\mu_0}$ is an $\rE(3)$-coadjoint orbit, as well.

%%%%%%%%%%%%%%%%%%%%%%%%%%%%%%%%%%%%%%%%%%%%%%%%%%%%%%%%%%%%%%%%%%%%%%%%%%%%%%%%%%%%
\subsection{The Fermat equations}
%%%%%%%%%%%%%%%%%%%%%%%%%%%%%%%%%%%%%%%%%%%%%%%%%%%%%%%%%%%%%%%%%%%%%%%%%%%%%%%%%%%%

In order to describe light rays in a refractive medium of index $n\in{}C^1(\bbR^3,\bbR^{>0})$, let us modify the $1$-form (\ref{varpi0}) according to Fermat's prescription, $\varpi\leadsto{}n\varpi$, and start with the new $1$-form
\begin{equation}
\varpi=p\,n(\bx)\la\bu,d\bx\ra
\label{varpi-n}
\end{equation}
on the bundle $ST\bbR^3$. 

\goodbreak

Again, the characteristic foliation of $\sigma=d\varpi$ should lead to ordinary differential equations governing light rays in such a medium. Indeed, $(\delta\bu,\delta\bx)\in\ker(\sigma)$ iff $\la\delta(n\bu),\delta'\bx\ra-\la\delta'(n)\bu+n\delta'\bu,\delta\bx\ra+\lambda\la\bu,\delta'\bu\ra=0$ for all $\delta'\bu,\delta'\bx\in\bbR^3$, where $\lambda$ is a real Lagrange multiplier. Redefining $\alpha=\lambda/n$, we get
\begin{equation}
(\delta\bu,\delta\bx)\in\ker(\sigma)
\Longleftrightarrow
\left\{
\begin{array}{rcll}
\delta(n\bu)&=&\alpha\,\grad\,n\\
\delta\bx&=&\alpha\,\bu
\end{array}
\right.
\label{kersigma-n}
\end{equation}
with $\alpha\in\bbR$, i.e., a $1$-dimensional foliation which yields---if we put $\alpha=\delta{t}$ where~$t$ is now arc-length---Fermat's equations of geometrical optics in an isotropic medium of refractive index~$n$. Notice that the system (\ref{kersigma-n}) is independent of the color $p$. 

%We will show below how the Fermat prescription stems from the more general context of Riemannian geometry.

%%%%%%%%%%%%%%%%%%%%%%%%%%%%%%%%%%%%%%%%%%%%%%%%%%%%%%%%%%%%%%%%%%%%%%%%%%%%%%%%%%%%
\subsection{The spinning and colored Euclidean coadjoint orbits}
%%%%%%%%%%%%%%%%%%%%%%%%%%%%%%%%%%%%%%%%%%%%%%%%%%%%%%%%%%%%%%%%%%%%%%%%%%%%%%%%%%%%

%-----------------------------------------------------------------------------------
\subsubsection{The manifold of circularly polarized light rays}
%-----------------------------------------------------------------------------------

Apart from the trivial coadjoint orbit and the $2$-spheres characterized by the invariants $C=0$ and $C''=s^2$ with $s>0$, there exists, most interestingly, another class of $\SE(3)$-coadjoint orbits defined by the invariants $C=p^2$, with $p>0$ (color), and $C'=s p$ where $s\neq0$ stands for spin. The orbit passing through $\mu_0=(\bell_0,\bp_0)$ where $\bell_0=(s,0,0)$ and $\bp_0=(p,0,0)$ is again $\cO_{\mu_0}\cong{}TS^2$ and is endowed with the symplectic structure coming from the $1$-form (\ref{canonical1form}) on the group $\SE(3)$ which reads
\begin{equation}
\varpi=p\la\bu,d\bx\ra-s\la\bv,d\bw\ra.
\label{varpisp}
\end{equation}

\goodbreak

This $1$-form is, by construction, $\SE(3)$-invariant. The associated momentum mapping, see \cite{Sou}, $J:\SE(3)\to\cO_{\mu_0}\subset\se(3)^*:(R,\bx)\mapsto(\bell,\bp)$ is actually given by $\varpi(\delta_Z{g})=J(g)\cdot{}Z$ where $g\mapsto\delta_Z{g}$ is the fundamental vector field associated with $Z\in\se(3)$. 
If $Z=(\bomega,\bgamma)$, we get $\delta_Z(\bu,\bv,\bw,\bx)=(\bomega\times\bu,\bomega\times\bv,\bomega\times\bw,\bomega\times\bx+\bgamma)$. We readily obtain $\varpi(\delta_Z{g})=p\la\bu,\bomega\times\bx+\bgamma\ra-s\la\bv,\bomega\times\bw\ra=\la\bell,\bomega\ra+\la\bp,\bgamma\ra$, hence
\begin{equation}
\left\{
\begin{array}{rcl}
\bell&=&\bx\times\bp+s\bu\\[8pt]
\bp&=&p\bu
\end{array}
\right.
\label{J}
\end{equation}
allowing us to interpret $\bell$ as the angular momentum, $s\bu$ as the spin (or polariza\-tion) vector and $\bp$ as the linear momentum of the light ray. We call helicity the sign of the spin invariant, $\chi=\sign(s)$.

\goodbreak

Note, \textit{en passant}, that the union of two $\SE(3)$-coadjoint orbits defined by the invariants $(p,s)$ and $(p,-s)$ is symplecto\-morphic to a single $\rE(3)$-coadjoint orbit.

An intermediate stage between these classical models and their geometrically quantized version is prequantization \cite{Sou}. Here, the latter construction would restrict spin to be half integral, $s\in\half\bbZ\hbar$. To describe spinning light rays, we will naturally put $s=\chi\hbar$. Euclidean co\-adjoint orbits with $\chi=+1$ (resp. $\chi=-1$) describe right-handed (resp. left-handed) circularly polarized light rays \cite{Sou,GS}. For the sake of completeness, let us mention that one often writes $p=k\hbar$ where $k$ may be interpreted as the (Euclidean) ``wave number'' of the light rays.

Straightforward computation yields $d\la\bv,d\bw\ra=\Surf$, i.e., the surface element of the $2$-sphere described by $\bu=\bv\times\bw$, namely $\Surf(\delta\bu,\delta'\bu)=\la\bu,\delta\bu\times\delta'\bu\ra$. The exterior derivative $\sigma=d\varpi$ takes on the form
\begin{equation}
\sigma(\delta{g},\delta'{g})=p\left[\la\delta\bu,\delta'\bx\ra-\la\delta'\bu,\delta\bx\ra\right]-s\la\bu,\delta\bu\times\delta'\bu\ra.
\label{sigmasp}
\end{equation}
for all $\delta{g},\delta'g\in{}T_g\,\SE(3)$; this $2$-form again descends to $ST\bbR^3$. The characteristic foliation of the latter $2$-form is, \textit{verbatim}, given by (\ref{kersigma0}): the spinning light rays in vacuum are nothing but the Euclidean geodesics. (As will be shown in the sequel, things will change dramatically for such light rays in a refractive medium.) 

The manifold $\cO_{\mu_0}\cong{}TS^2$ of spinning light rays is, just as before, parametrized by the pairs $\xi=(\bq,\bu)$ and endowed with the ``twisted'' symplectic $2$-form, $\omega$, viz.,
\begin{equation}
\sigma(\delta{g},\delta'{g})=\omega(\delta\xi,\delta'\xi)=p\left[\la\delta\bu,\delta'\bq\ra-\la\delta'\bu,\delta\bq\ra\right]
-s\la\bu,\delta\bu\times\delta'\bu\ra.
\label{omegasp}
\end{equation}

\goodbreak

%-----------------------------------------------------------------------------------
\subsubsection{A noncommutative wave plane}
%-----------------------------------------------------------------------------------

A new phenomenon then appears, which we link to the previously introduced twisted symplectic structure, namely noncommutativity of each fiber $T_\bu{}S^2$ viewed as the ``wave plane''~$\bu^\perp$ ortho\-gonal to the direction $\bu$ of the ray. 
Indeed, let us define linear coordinates $q_i=\la\bv_i,\bq\ra$ with $\bv_i\in\bbR^3$ for $i=1,2$ in that plane. Straightforward computation of the Poisson bracket of $q_1$ and $q_2$, with respect to the symplectic structure $\omega$ in (\ref{omegasp}), yields
$
\{q_1,q_2\}=-\omega^{-1}(dq_1,dq_2)=(1/p^2)\la\bell,\bv_1\times\bv_2\ra
$
where the angular momentum $\bell$ is as in (\ref{J}). If $\bv_1$ and $\bv_2$ form an orthonormal basis of~$\bu^\perp$  such that $\bv_1\times\bv_2=\bu$, then
\begin{equation}
\{q_1,q_2\}=\frac{s}{p^2}
\label{NC}
\end{equation}
implying noncommutativity of each (Lagrangian) space $\bu^\perp$ consisting of all rays parallel to the direction $\bu$.

%-----------------------------------------------------------------------------------
\subsubsection{Euclidean polarized light rays as stationary massless states of the Poincar\'e and Galilei groups}
%-----------------------------------------------------------------------------------

Let us discuss here, for the sake of completeness, the relationship between this purely Euclidean model of spinning and colored light rays and the classical models of massless particles pertaining to either the relativistic or the non\-relativistic framework.  

$\bullet$ The spin-$s$, massless, coadjoint orbit of $\rE(3,1)_0=\rO(3,1)_0\ltimes\bbR^4$, the connected Poincar\'e group, has the topology of the cotangent bundle of the punctured light cone, described by $(\bx,\bp)\in{}T(\bbR^3\!\setminus\!\{0\})$, and is endowed with the symplectic $2$-form~$\widetilde{\omega}$ given, in a Lorentz frame, by
\begin{equation}
\widetilde{\omega}(\delta(\bx,\bp),\delta'(\bx,\bp))=\la\delta\bp,\delta'\bq\ra-\la\delta'\bp,\delta\bq\ra-\frac{s}{\Vert\bp\Vert^3}\la\bp,\delta\bp\times\delta'\bp\ra
\label{omegasPoincare}
\end{equation}
corresponding to Equation (17.145) in \cite{Sou}; see also \cite{DET,DE}.
The Lie subgroup $(\bbR,+)$ of time translations acts in a Hamiltonian way on it. Its infinitesimal action is given by $\delta_\varepsilon(\bx,\bp)=\varepsilon(-c\bp/\Vert\bp\Vert,0)$ where $\varepsilon\in\bbR$ and $c$ denotes the velocity of light in vacuum. The associated moment map is the energy $E:(\bx,\bp)\mapsto{}c\Vert\bp\Vert$, namely $\widetilde{\omega}(\delta_\varepsilon(\bx,\bp))=d(E\varepsilon)$. 

\goodbreak

The submanifold $\imath_E:ST\bbR^3\hookrightarrow{}T(\bbR^3\!\setminus\!\{0\})$ defined by a fixed positive energy
\begin{equation}
E=c p
\label{E=p}
\end{equation}
where the constant $p>0$ stands for ``color'', is indeed the pre\-symplectic manifold $(ST\bbR^3,\sigma)$ given by (\ref{sigmasp}), where $\bu$ is the direction of $\bp$. This presymplectic structure is preserved by the group $\rA(3,1)_0=\SE(3)\times\bbR\subset\rE(3,1)_0$. (The ``Aristotle group'', $\rA(3,1)$, is the centralizer of time translations in the Poincar\'e group $\rE(3,1)$.) The Marsden-Weinstein (MW) reduced symplectic manifold $T(\bbR^3\!\setminus\!\{0\})//(\bbR,+)$, i.e., the stationary relativistic states of energy (\ref{E=p}), is then symplectomorphic to our coadjoint orbit $(TS^2,\omega)$ of the Euclidean group $\SE(3)=\rA(3,1)_0/\bbR$, viz.,
\begin{equation}
\imath_E^*\,\widetilde{\omega}=\widetilde{\pi}^*\omega
\label{omegasPoincareReduced}
\end{equation}
where $\widetilde{\pi}:(\bx,\bp)\mapsto(\bq=\bx-\bu\la\bu,\bx\ra,\bu=\bp/\Vert\bp\Vert)$, and $\omega$ is as in (\ref{omegasp}).

$\bullet$ The (massless) coadjoint orbit of $\Gal(3,1)_0=\SE(3)\ltimes\bbR^4$, the connected Galilei group, with Casimir invariants $p$ (color) and $s$ (spin) is plainly symplecto\-morphic to $(TS^2\times{}T\bbR,\widehat{\omega})$ described by the quadruples $(\bq,\bu,t,E)$, where %(slight abuse of notation)
\begin{equation}
\widehat{\omega}=\omega-dE\wedge{}dt
\label{omegaspGal}
\end{equation}
and $\omega$ is given by (\ref{omegasp}); cf. Proposition (14.53) of \cite{Sou}. Time translations, $(\bbR,+)$, act in an Hamiltonian way via $(\bq,\bu,t,E)\mapsto(\bq,\bu,t+e,E)$ where $e\in\bbR$. The associated moment map is clearly given by the energy $(\bq,\bu,t,E)\mapsto{}E$.

\goodbreak

The submanifold $\jmath_E:TS^2\times\bbR\hookrightarrow{}TS^2\times{}T\bbR^+$ of Galilei massless states of fixed energy
\begin{equation}
E=\const
\label{E}
\end{equation}
is endowed with the presymplectic $2$-form $\jmath_E^*\,\widehat{\omega}$. Again, its structure is preserved by~$A(3,1)_0$. (Note that $A(3,1)=E(3,1)\cap\Gal(3,1)\subset\GL(5,\bbR)$.) The MW-reduced symplectic manifold $(TS^2\times{}T\bbR)//(\bbR,+)$, i.e., the non\-relativistic stationary states of energy (\ref{E}), is, again, symplectomorphic to our $\SE(3)$-coadjoint orbit, viz.,
\begin{equation}
\jmath_E^*\,\widehat{\omega}=\widehat{\pi}^*\omega
\label{omegasGalileiReduced}
\end{equation}
where $\widehat{\pi}:(\bq,\bu,t)\mapsto(\bq,\bu)$, and $\omega$ is as in (\ref{omegasp}).

We have, hence, shown that the connected symplectic manifolds of Euclidean spinning light rays, naturally arise as the MW-reduced manifolds of stationary states of given energy, $E=\const$, for the 
massless spinning coadjoint orbits of the connected Poincar\'e---as well as Galilei---group.\footnote{The two-component $E(3)$-coadjoint orbits are reduced MW-manifolds of the orthochronous Poincar\'e and Galilei groups.}

%%%%%%%%%%%%%%%%%%%%%%%%%%%%%%%%%%%%%%%%%%%%%%%%%%%%%%%%%%%%%%%%%%%%%%%%%%%%%%%%%%%%
%%%%%%%%%%%%%%%%%%%%%%%%%%%%%%%%%%%%%%%%%%%%%%%%%%%%%%%%%%%%%%%%%%%%%%%%%%%%%%%%%%%%
\section{A general framework for spinoptics}\label{SecSpinOptics}
%%%%%%%%%%%%%%%%%%%%%%%%%%%%%%%%%%%%%%%%%%%%%%%%%%%%%%%%%%%%%%%%%%%%%%%%%%%%%%%%%%%%
%%%%%%%%%%%%%%%%%%%%%%%%%%%%%%%%%%%%%%%%%%%%%%%%%%%%%%%%%%%%%%%%%%%%%%%%%%%%%%%%%%%%

%%%%%%%%%%%%%%%%%%%%%%%%%%%%%%%%%%%%%%%%%%%%%%%%%%%%%%%%%%%%%%%%%%%%%%%%%%%%%%%%%%%%
\subsection{The Fermat prescription revisited}
%%%%%%%%%%%%%%%%%%%%%%%%%%%%%%%%%%%%%%%%%%%%%%%%%%%%%%%%%%%%%%%%%%%%%%%%%%%%%%%%%%%%

Let us recall how the Fermat equations describing light propagation in an isotropic and inhomogeneous medium can be interpreted as those of the geodesics of a metric conformally related to the flat spatial Euclidean metric; see also \cite{CN,CN2}.

\goodbreak

Returning to the expression (\ref{varpi-n}) which we rewrite as $\varpi=p\,n(\bx)^2\la{}n(\bx)^{-1}\bu,d\bx\ra$, we introduce the new, curved, metric $\rg=n^2\la\,\cdot\,,\,\cdot\,\ra$ on $\bbR^3$ and the $\rg$-unitary vector $U=n^{-1}\bu$; we also put $X=\bx$ to keep the notation coherent. With these preparations, $(M\cong\bbR^3,\rg)$ becomes a Riemannian $3$-manifold while its spherical tangent bundle $STM=\{\eta=(U,X)\in{}TM\,\vert\,\rg(U,U)=1\}$ becomes endowed with the $1$-form
\begin{equation}
\varpi=p\,\rg(U,dX)
\label{varpiSTM}
\end{equation}
which (up to an overall constant factor $p$) stems from the canonical $1$-form of~$T^*M$ via the metric~$\rg$. 

\goodbreak

We already know that the characteristic foliation of $\sigma=d\varpi$ yields the Fermat equations, see (\ref{kersigma-n}). Let us briefly recall why this foliation also provide us with the geodesic flow associated with the Fermat metric $\rg=\rg_{ij}(X)\,dX^i\otimes{}dX^j$, where
\begin{equation}
\rg_{ij}(X)=n(X)^2\delta_{ij}
\label{FermatMetric}
\end{equation}
with $i,j=1,2,3$.
(Note that the metric $\rg=n_i(X)^2\delta_{ij}\,dX^i\otimes{}dX^j$ would readily enable us to deal with some special anisotropic media \cite{CN}.)
Call $\nabla$ the Levi-Civita connection of~$(M,\rg)$ and~$\Gamma_{ij}^k$ its local components. We will denote by $d^\nabla$ the exterior covariant derivative of tensor fields, e.g., locally, $d^\nabla{}U^k=dU^k+\Gamma_{ij}^k dX^i{}U^j$. Then, the $2$-form $\sigma=d\varpi=p\,\rg_{ij}(X)\,d^\nabla{}U^i\wedge{}dX^j$ writes alternatively
\begin{equation}
\sigma(\delta\eta,\delta'\eta)
=
p\left[\rg(\delta^\nabla{}U,\delta'X)-\rg(\delta'^\nabla{}U,\delta{}X)\right]
\label{sigmaSTM}
\end{equation}
for all $\delta\eta,\delta'\eta\in{}T_\eta\,STM$. Its characteristic foliation is integrated by the geodesic flow. Actually, $\delta(U,X)\in\ker(\sigma)$ iff $\rg(\delta^\nabla{U},\delta'X)-\rg(\delta'^\nabla{}U,\delta{X})+\lambda\rg(U,\delta'^\nabla{}U)=0$ for all $\delta'^\nabla{}U,\delta'X\in{}TM$, where $\lambda\in\bbR$. We hence obtain
\begin{equation}
\delta(U,X)\in\ker(\sigma)
\Longleftrightarrow
\left\{
\begin{array}{rcll}
\delta^\nabla{}U&=&0\\
\delta{X}&=&\lambda{}U
\end{array}
\right.
\label{kersigmaSTM}
\end{equation}
with $\lambda\in\bbR$, which we recognize as the geodesic foliation for $(M,\rg)$.

\goodbreak

%%%%%%%%%%%%%%%%%%%%%%%%%%%%%%%%%%%%%%%%%%%%%%%%%%%%%%%%%%%%%%%%%%%%%%%%%%%%%%%%%%%%
\subsection{Spinoptics \& minimal coupling to a Fermat metric}
%%%%%%%%%%%%%%%%%%%%%%%%%%%%%%%%%%%%%%%%%%%%%%%%%%%%%%%%%%%%%%%%%%%%%%%%%%%%%%%%%%%%

Let us now tackle the geometric description of spinning light rays in an arbitrary Rieman\-nian, orientable, $3$-manifold $(M,\rg,\vol_\rg)$. As we will see, this novel approach allowing for a complete treatment of the geodesic deviation of spinning light rays in a generalized Fermat metric borrows much from general relativity, namely from the Papapetrou-Dixon-Souriau equations of motion of test particles in the gravitational field, see \cite{Pap,Dix,Sou2} and, e.g., \cite{DFS,Kun,Duv,Ste}.

\goodbreak

%-----------------------------------------------------------------------------------
\subsubsection{Minimal coupling}
%-----------------------------------------------------------------------------------

The procedure involved is already known as ``minimal coupling'' to a classical external field. In our context, it will simply consist in considering, instead of the Euclidean group (viewed as the trivial $\SO(3)$-principal bundle over $E^3$), the bundle, $\SO(M)\to{}M$, of oriented orthonormal frames of $M$ and in the replacement
\begin{equation}
\la\,\cdot\,,\,\cdot\,\ra\leadsto\rg
\qquad
\&
\qquad
d\leadsto{}d^\nabla
\label{defMinimalCoupling}
\end{equation}
into the $1$-form (\ref{varpisp}) with invariants $p$ and $s$. With the above notation, this enables us to generalize (\ref{varpiSTM}) with the following $1$-form
\begin{equation}
\varpi=p\,\rg(U,dX)-s\,\rg(V,d^\nabla{}W)
\label{varpispMg}
\end{equation}
on the $\SO(3)$-principal bundle $\SO(M)\to{}M$ described by $g=((U,V,W),X)$ where $(U,V,W)$ is a $\rg$-orthonormal basis of $T_XM$ such that $\vol_\rg(U,V,W)=1$.

\goodbreak

Taking advantage of the above observation that the characteristic foliation of $\sigma=d\varpi$ yields the equations of light rays, we contend that the equations of ``geometrical spinoptics'' based on $(M,\rg)$, for color $p$ and spin $s=\chi\hbar$ (with~$\chi=\pm1$), are associated with the foliation $\ker(\sigma)$ we are now ready to determine explicitly.

%-----------------------------------------------------------------------------------
\subsubsection{Notation and miscellaneous formul\ae}
%-----------------------------------------------------------------------------------

We denote by $j(\delta'{X})$ the $\rg$-skew symmetric (cross-product) operator of $T_XM$ defined by $\rg(\delta{X},j(\delta'{X})\delta''X)=\vol_\rg(\delta{X},\delta'{X},\delta''X)$. Putting, e.g., $\overline{V}=\rg(V)=\rg(V,\,\cdot\,)$, we have
\begin{equation}
U=j(V,W)
\qquad
\Longleftrightarrow
\qquad
j(U)=W\overline{V}-V\overline{W}.
\label{j}
\end{equation}

\goodbreak

The curvature, $R$, of the Levi-Civita connection $\nabla$ of $(M,\rg)$ is defined by \begin{equation}
R(\delta{X},\delta'X)\delta''X\equiv\delta^\nabla\delta'^\nabla\delta''X-\delta'^\nabla\delta^\nabla\delta''X-[\delta,\delta']^\nabla\delta''X
\label{Riemann}
\end{equation}
where $X\mapsto[\delta,\delta']X$ is the Lie bracket of the vector fields $X\mapsto\delta{X}$ and $X\mapsto\delta'{X}$. Its local expression is given by $R^\ell_{ijk}\,\partial_\ell=\nabla_i\nabla_j\partial_k-\nabla_j\nabla_i\partial_k$, where $\partial_k=\partial/\partial{}X^k$, for $i,j,k,\ell=1,\ldots,3$. 

The Ricci tensor $\Ric(\delta{X},\delta'{X})\equiv\Tr(\delta''X\mapsto{}R(\delta''{X},\delta{X})\delta'X)$ has local expression $R_{jk}=R^i_{ijk}$. 

\goodbreak

Now, since $\dim(M)=3$, we have 
\begin{equation}
R_{ijk}^\ell
=
-\left(
R_{ik}\,\delta_j^\ell-R_{jk}\,\delta_i^\ell+R_j^\ell\,\rg_{ik}-R_i^\ell\,\rg_{jk}\right)%\\
%&&
+\half\,R\left(\rg_{ik}\,\delta_j^\ell-\rg_{jk}\,\delta_i^\ell\right)
\label{Riemann3D}
\end{equation}
where $R=R_{ij}\rg^{ij}$ is the scalar curvature.

If $\Omega=j(U)$ is as in (\ref{j}), we find
\begin{equation}
\rg(V,R(\delta{X},\delta'X)W)\equiv\half\rg(\delta{X},R(\Omega)\delta'X)
\label{defROmega}
\end{equation}
 where
the operator~$R(\Omega)$ is given, via (\ref{Riemann3D}), by the $\rg$-skew symmetric operator
\begin{equation}
R(\Omega)=-2(\Ric\,\Omega+\Omega\,\Ric)+R\,\Omega.
\label{ROmega}
\end{equation}
The scalar function $R(\Omega,\Omega)=-\Tr(R(\Omega)\Omega)$ will also be needed; from (\ref{ROmega}) and $\Omega^2=U\overline{U}-\bone$, we get the remarkable expression
\begin{equation}
\frac{1}{4}R(\Omega,\Omega)=\Ein(U,U)
\label{ROmegaOmega}
\end{equation}
where $\Ein=\Ric -\half{}R\,\rg$ is the Einstein tensor of the metric $\rg$.

For the conformally flat metric (\ref{FermatMetric}), we readily find the Christoffel symbols
\begin{equation}
\Gamma_{ij}^k=\frac{1}{n}\left(
\partial_i{n}\,\delta_j^k+\partial_j{n}\,\delta_i^k-\partial_\ell{n}\,\delta^{k\ell}\delta_{ij}
\right),
\label{ChristoffelFermat}
\end{equation}
the Ricci tensor
\begin{equation}
R_{ij}
=
\frac{2}{n^2}\partial_in\,\partial_jn
-
\frac{1}{n}\partial_i\partial_jn
-\frac{1}{n}\Delta{n}\,\delta_{ij},
\label{RicciFermat}
\end{equation}
where $\Delta{n}=\delta^{ij}\partial_i\partial_jn$, and the scalar curvature
\begin{equation}
R
=
\frac{2}{n^4}\Vert{}dn\Vert^2
-\frac{4}{n^3}\Delta{n}.
\label{ScalarCurvatureFermat}
\end{equation}
%%% At last, we find the Einstein tensor to be
%%% \begin{equation}
%%% E_{ij}
%%% =
%%% \frac{2}{n^2}\partial_in\,\partial_jn
%%% -
%%% \frac{1}{n}\partial_i\partial_jn
%%% +\left[\frac{1}{n}\Delta{n}-\frac{1}{n^2}\Vert{}dn\Vert^2
%%% \right]\delta_{ij}.
%%% \label{EinsteinFermat}
%%% \end{equation}

%Defining the ``velocity'' of light as ${v}=1/n$, we get the alternative expressions
%\begin{equation}
%R_{ij}
%=
%\frac{1}{{v}}\partial_i\partial_j{v}
%+
%\left[\frac{1}{{v}}\Delta{{v}}-\frac{2}{{v}^2}\Vert{}d{v}\Vert^2\right]\delta_{ij},
%\label{RicciFermatBis}
%\end{equation}
%and
%\begin{equation}
%R
%=
%4{v}\,\Delta{v}
%-
%6\Vert{}d{v}\Vert^2,
%\label{ScalarCurvatureFermatBis}
%\end{equation}
%and
%\begin{equation}
%E_{ij}
%=
%\frac{1}{{v}}\partial_i\partial_j{v}
%-
%\left[\frac{1}{{v}}\Delta{{v}}-\frac{1}{{v}^2}\Vert{}d{v}\Vert^2\right]\delta_{ij}.
%\label{EinsteinFermatBis}
%\end{equation}

%-----------------------------------------------------------------------------------
\subsubsection{The general system for spinoptics}
%-----------------------------------------------------------------------------------

Let us work out the expression of the $2$-form $\sigma=d\varpi$ on $\SO(M)$ where $\varpi$ is as in~(\ref{varpispMg}). 

\goodbreak

We find, remembering~(\ref{Riemann}),
 \begin{eqnarray*}
\sigma(\delta{g},\delta'{g})
&=&
\delta(\varpi(\delta'{g}))-\delta'(\varpi(\delta{g}))-\varpi([\delta,\delta']{g}))\\
&=&
p\left[\rg(\delta^\nabla{}U,\delta'X)-\rg(\delta'^\nabla{}U,\delta{X})\right]\\
&&
-s\left[\rg(\delta^\nabla{}V,\delta'^\nabla{}W)-\rg(\delta'^\nabla{}V,\delta^\nabla{}W)
-\rg(V,R(\delta{X},\delta'X)W)
\right]
\end{eqnarray*}
for all $\delta{g},\delta'{g}\in{}T_g\,\SO(M)$. 
Now, using the closure formula $U\overline{U}+V\overline{V}+W\overline{W}=\bone$, we readily get $\rg(\delta^\nabla{}V,\delta'^\nabla{}W)=\rg(\delta^\nabla{}V,[U\overline{U}+V\overline{V}+W\overline{W}]\delta'^\nabla{}W)=\rg(\delta^\nabla{}V,U\overline{U}\delta'^\nabla{}W)=\rg(\delta^\nabla{}U,V\overline{W}\delta'^\nabla{}U)$ since $(U,V,W)$ is a $\rg$-orthogonal frame. We then deduce that $\rg(\delta^\nabla{}V,\delta'^\nabla{}W)-\rg(\delta'^\nabla{}V,\delta^\nabla{}W)=\rg(\delta^\nabla{}U,[V\overline{W}-W\overline{V}]\delta'^\nabla{}U)=-\rg(\delta^\nabla{}U,j(U)\delta'^\nabla{}U)=\vol_\rg(U,\delta^\nabla{}U,\delta'^\nabla{}U)$ in view of~(\ref{j}). At last, we get
\begin{eqnarray}
\sigma(\delta{g},\delta'{g})
\nonumber
&=&
p\left[\rg(\delta^\nabla{}U,\delta'X)-\rg(\delta'^\nabla{}U,\delta{X})\right]\\
\label{sigmaMg}
&&
-\half\,s\,\rg(\delta{X},R(\Omega)\delta'{X})\\
\nonumber
&&
-s\,\vol_\rg(U,\delta^\nabla{}U,\delta'^\nabla{}U)
\end{eqnarray}
with the help of (\ref{defROmega}) and the shorthand notation $\Omega=j(U)$. The $2$-form (\ref{sigmaMg}) turns out to descend, again, to the spherical tangent bundle~$STM$, described by $\eta=(U,X)$. Still denoting $\sigma$ that $2$-form on $STM$, we compute its kernel.

Introducing, just as before, a real Lagrange multiplier, $\lambda$, for the constraint $\rg(U,U)=1$, we readily find that $\delta(U,X)\in\ker(\sigma)$ iff $p\delta^\nabla{}U+\half\,s{}R(\Omega)\delta{X}=0$ and $p\delta{X}+s{}j(U)\delta^\nabla{}U=\lambda{}U$. This entails $p\delta{X}-s^2/(2p)\Omega{}R(\Omega)\delta{X}=\lambda{}U$, prompting the Ansatz $\delta{X}=\alpha{}U+\beta{}s^2\Omega{}R(\Omega)U$, for some $\alpha,\beta\in\bbR$ still to be determined. We thus get
$p^2(\alpha{}U+\beta\Omega{}R(\Omega)U)-\half\,s^2\alpha{}\Omega{}R(\Omega)U
-\half\,s^4\beta{}\Omega{}\left[R(\Omega)\Omega{}R(\Omega)\right]U=p\lambda{}U$.
Recall that, if $A,B$ are $\rg$-skew symmetric operators of $T_XM$, then $ABA=\half\Tr(AB)A$. This enables us to compute the above bracketed term, viz., $R(\Omega)\Omega{}R(\Omega)=-\half{}R(\Omega,\Omega)R(\Omega)$, and to find $\alpha=\lambda/p$ and $\beta=\lambda/(2p(p^2+\frac{1}{4}\,s^2{}R(\Omega,\Omega))$. 
Invoking (\ref{ROmegaOmega}), we end up with the spinoptics system governing the trajectories of spinning light rays in a Riemannian $3$-manifold~$(M,\rg)$, viz.,
\begin{equation}
\delta(U,X)\in\ker(\sigma)
\Longleftrightarrow
\left\{
\begin{array}{rcll}
p\,\delta^\nabla{}U
&=&
%\displaystyle
-\half\,s{}R(\Omega)\delta{X}\\[10pt]
%\delta{X}&=&\alpha\left[
%\displaystyle
%U+\frac{s^2}{2p^2+\half{}s^2{}R(\Omega,\Omega))}j(U)R(j(U))U
%\right]
\delta{X}
&=&
\alpha\left[
\displaystyle
U+\frac{s^2\,\Omega\,R(\Omega)U}{2\left[p^2+ s^2\Ein(U,U)\right]}
\right]
\end{array}
\right.
\label{kersigmaSTMg}
\end{equation}
with $\Omega=j(U)$, and $\alpha\in\bbR$.

\goodbreak

In the $(3+1)$-dimensional setting of general relativity, a similar system (the Papapetrou-Dixon-Souriau equations) would describe geodesic deviation and spin precession of spinning test particles in a gravitational background field. 

We stress that, in spinoptics (as well as in general relativity \cite{Sou2}), ``velocity'' $\delta{X}$ and ``momentum'' $U$ fail to be parallel (see (\ref{kersigmaSTMg})). We will see how this phenomenon gives rise to subtle physical effects such as the Optical Hall Effect.

%-----------------------------------------------------------------------------------
\subsubsection{Fermat spinoptics}
%-----------------------------------------------------------------------------------

As an illustration of the preceding results, let us write the exact spinoptics equations specialized to the Fermat metric~(\ref{FermatMetric}) associated with a refractive index $n$. Just as before, we denote by $p(=k\hbar)$ the color and $s(=\chi\hbar)$ the spin of the model. 
Although this could be deduced from the very general system (\ref{kersigmaSTMg}), we choose to simply start from the $1$-form (\ref{varpispMg}) as this procedure actually yields all parameters adapted to the model in a straightforward fashion.

\goodbreak

We begin here with $(M,\rg)=(\bbR^3,n^2\la\,\cdot\,,\,\cdot\,\ra)$ where $n\in{}C^2(\bbR^3,\bbR^{>0})$, and, upon defining the Euclidean frame $(\bu,\bv,\bw)=(nU,nV,nW)$, express the $1$-form (\ref{varpispMg}) as a new $1$-form on~$\SE(3)$ which is parametrized by $g=((\bu,\bv,\bw),\bx)$ as in Section~\ref{SecEuclideanOptics}. Using the expression (\ref{ChristoffelFermat}) of the Christoffel symbols, we readily notice that $\rg(V,d^\nabla{}W)=\la\bv,d\bw\ra+n(\bx)^{-1}\left[\la\bv,d\bx\ra\,dn(\bw)-\la\bw,d\bx\ra\,dn(\bv)\right]$. Introducing now the ``velocity''
\begin{equation}
v=\frac{1}{n}
\label{velocity}
\end{equation}
and its gradient
\begin{equation}
\bg=\grad(v),
\label{g}
\end{equation}
we find the new spin term $\rg(V,d^\nabla{}W)=\la\bv,d\bw\ra-n\left[\la\bv,d\bx\ra\,\la\bw,\bg\ra-\la\bw,d\bx\ra\,\la\bv,\bg\ra\right]=\la\bv,d\bw\ra+n\la(\bv\times\bw)\times\bg,d\bx\ra$. 

\goodbreak

This entails, that $\varpi=p\,\rg(U,dX)-s\,\rg(V,d^\nabla{}W)$ retains the following form
\begin{equation}
\varpi=\la\wbp,d\bx\ra-s\la\bv,d\bw\ra
\label{varpiSpinFermat}
\end{equation}
where
\begin{equation}
\wbp=n(\bx)(p\bu+s\bg\times\bu)
\label{p}
\end{equation}
can be consistently interpreted as the spin-dependent ``momentum'' of the system.

\goodbreak

Easy computation gives the exterior derivative $\sigma=d\varpi$ of the $1$-form (\ref{varpiSpinFermat}), namely
\begin{equation}
\sigma(\delta{g},\delta'{g})=\la\delta\wbp,\delta'\bx\ra-\la\delta'\wbp,\delta\bx\ra-s\la\bu,\delta\bu\times\delta'\bu\ra.
\label{sigmaSpinFermat}
\end{equation}
for all $\delta{g},\delta'g\in{}T_g\,\SE(3)$, see (\ref{sigmasp}). This $2$-form descends to $ST\bbR^3$ and has, generically, rank $4$; computing its kernel needs some more effort.  

Let us denote by
\begin{equation}
\nabla\bg=\frac{\partial\bg}{\partial\bx}
\label{nablag}
\end{equation}
the (symmetric) second derivative of the velocity $v$.

We find, using (\ref{p}), and in the same way as before, $(\delta\bu,\delta\bx)\in\ker(\sigma)$ iff $\delta\wbp+n\bg\la\wbp,\delta\bx\ra-ns(\nabla\bg)\bu\times\delta\bx=0$ and $pn\delta\bx-ns\bg\times\delta\bx+s\bu\times\delta\bu=\lambda\bu$ where $\lambda\in\bbR$ is a Lagrange multiplier. Taking the cross-product of the latter equation by $\bu$ yields $s\delta\bu=\bu\times{}n(p\delta\bx-s\bg\times\delta\bx)$, which can be inserted into the former equation with the help of (\ref{p}). In doing so, using the following partial result $\delta\bu+(s/p)\bg\times\delta\bu=\left[(np/s)\bu+n\bg\times\bu+(sn/p)\bg\la\bg,\bu\ra\right]\times\delta\bx$, we end up with 
\begin{equation}
\frac{p^2n^2}{s}\bu\times\delta\bx+n^2s\Vert\bg\Vert^2\bu\times\delta\bx
-
sn\Big[j(\bu)\,\nabla\bg+\nabla\bg\,j(\bu)\Big]\delta\bx
=0
\label{intermediateEq}
\end{equation}
where, see (\ref{j}), $j(\bu):\delta\bx\mapsto\bu\times\delta\bx$ is the Euclidean cross-product operator. We still need to compute the last anticommutator in the above equation; it is given by the general formula $j(\bu)\,H+H\,j(\bu)=j(-H\bu+\Tr(H)\bu)$ for any $\bu\in\bbR^3$ and any symmetric $H\in{}L(\bbR^3)$. Equation (\ref{intermediateEq}) reduces then to
\begin{equation}
\delta\bx
= 
\beta
\left[
\frac{p^2n^2}{s}\bu+n^2s\Vert\bg\Vert^2\bu
+
sn(\nabla\bg)\bu
-
sn\,\Tr(\nabla\bg)\bu
\right]
\label{deltaU}
\end{equation}
for some $\beta\in\bbR$.

We can finally write the system defining the kernel of our $2$-form as
\begin{equation}
\begin{array}{c}
(\delta\bu,\delta\bx)\in\ker(\sigma)\\[10pt]
%\\
%\Longleftrightarrow
\Updownarrow\\[10pt]
%\\
\left\{
\begin{array}{rcll}
s\delta\bu
&=&
\displaystyle
\frac{p}{v}\,\bu\times\left(\bone-\frac{s}{p}j(\bg)\right)\delta\bx\\[12pt]
\delta\bx
&=&
\alpha\left[
\displaystyle
a\bu
+
\frac{v{}s^2}{p^2}\,\nabla_\bu\,\bg
\right]
\end{array}
\right.
\end{array}
\label{kersigmaSpinFermat}
\end{equation}
with $\alpha\in\bbR$, and where
\begin{equation}
a
=
1+\frac{s^2}{p^2}\Vert\bg\Vert^2-\frac{v{}s^2}{p^2}\rdiv(\bg)
\left[=
1+\frac{s^2}{p^2}\Vert\grad\,v\Vert^2-\frac{s^2}{p^2}v\Delta{v}\right].
\label{CoeffKersigmaSpinFermat}
\end{equation}
We have, equivalently\footnote{
Note that 
$\left(\bone+j(\bz)\right)^{-1}=(1+\Vert\bz\Vert^2)^{-1}\left(\bone-j(\bz)+\bz\overline{\bz}\right)$,
where $\overline{\bz}=\la\bz,\,\cdot\,\ra$, for all $\bz\in\bbR^3$.}
\begin{equation}
\begin{array}{c}
(\delta\wbp,\delta\bx)\in\ker(\sigma)\\[10pt]
%\\
%\Longleftrightarrow
\Updownarrow\\[10pt]
%\\
\left\{
\begin{array}{rcll}
\delta\wbp
&=&
\displaystyle
-\frac{1}{{v}}\la\wbp,\delta\bx\ra\,\bg
+\frac{s}{p}\frac{\partial\bg}{\partial\bx}
\left[\left(\bone+\frac{s}{p}j(\bg)\right)^{-1}\wbp\right]\times\delta\bx
\\[16pt]
\delta\bx
&=&
\alpha\left[
\displaystyle
a\bone
+
\frac{{v}{}s^2}{p^2}\,\frac{\partial\bg}{\partial\bx}
\right]
\displaystyle
\left(\bone+\frac{s}{p}j(\bg)\right)^{-1}\wbp
\end{array}
\right.
\end{array}
\label{kersigmaSpinFermatBis}
\end{equation}
with $\alpha\in\bbR$, in terms of the natural variables, viz., momentum $\wbp$ and position $\bx$. 

\goodbreak

This rather complicated system defining the equations for the trajectories of spinning light rays with color $p$ and spin $s$ in a refractive medium of index $n=1/{v}$ constitutes the novel differential equations of Fermat spinoptics, up to reparametrization. 

We clearly recover from (\ref{kersigmaSpinFermatBis}) the original Fermat equations (\ref{kersigma-n}) in the spinless case, $s=0$.

\goodbreak

Let us finish this Section by highlighting the relationship of our foliation for Fermat spinoptics to recent work of Onoda, Murakami and Nagaosa \cite{OMN}. Neglecting, in our system (\ref{kersigmaSpinFermatBis}), all terms involving second derivatives $\nabla\bg$ of the velocity~${v}$, and all quadratic terms $\Vert\bg\Vert^2$, we end up (choosing a parameter defined by $\alpha=1$), with the system
\begin{equation}
\left\{
\begin{array}{rcl}
\delta\wbp
&\cong&
\displaystyle
-\frac{1}{{v}}\la\wbp,\delta\bx\ra\,\bg
\\[12pt]
\delta\bx
&\cong&
\wbp
-
\displaystyle
\frac{s}{p}\,\bg\times\wbp
\end{array}
\right.
\label{EOMOMN}
\end{equation}
which, up to notation and reparametrization, exactly matches the first two Equations Of Motion put forward in~\cite{OMN}. The EOM (\ref{EOMOMN}) provide, hence, a linearization of our system~(\ref{kersigmaSpinFermatBis}) around~$\bg=0$.

%%%%%%%%%%%%%%%%%%%%%%%%%%%%%%%%%%%%%%%%%%%%%%%%%%%%%%%%%%%%%%%%%%%%%%%%%%%%%%%%%%%%
%%%%%%%%%%%%%%%%%%%%%%%%%%%%%%%%%%%%%%%%%%%%%%%%%%%%%%%%%%%%%%%%%%%%%%%%%%%%%%%%%%%%
\section{Spin Snell-Descartes' laws \& Optical Hall Effect}\label{SecSD-OHE}
%%%%%%%%%%%%%%%%%%%%%%%%%%%%%%%%%%%%%%%%%%%%%%%%%%%%%%%%%%%%%%%%%%%%%%%%%%%%%%%%%%%%
%%%%%%%%%%%%%%%%%%%%%%%%%%%%%%%%%%%%%%%%%%%%%%%%%%%%%%%%%%%%%%%%%%%%%%%%%%%%%%%%%%%%

We will, as a first test of our approach, establish the spinoptics version of the Snell-Descartes laws generalizing those of plain geometrical optics.

Consider the simplest case of a planar interface separating space into two regions $(M_1,n_1)$, resp. $(M_2,n_2)$, where $M_1=\{\bx\in\bbR^3\,\vert\,\la\bn,\bx\ra<0\}$ has refractive index $n_1=\const$, resp. $M_2=\{\bx\in\bbR^3\,\vert\,\la\bn,\bx\ra>0\}$ has refractive index $n_2=\const$, where $\bn$ is a unit vector, orthogonal to the interface (and pointing toward $M_2$), characterizing the optical device. 

Wishing to describe the laws of reflection and refraction, in geometrical terms, namely the scattering of spinning light rays by this device, we will resort to a theory developed by Souriau, namely ``symplectic scattering'' \cite{Sou}. See also \cite{GS2} for some further developments.

\goodbreak

%%%%%%%%%%%%%%%%%%%%%%%%%%%%%%%%%%%%%%%%%%%%%%%%%%%%%%%%%%%%%%%%%%%%%%%%%%%%%%%%%%%%
\subsection{Symplectic scattering}
%%%%%%%%%%%%%%%%%%%%%%%%%%%%%%%%%%%%%%%%%%%%%%%%%%%%%%%%%%%%%%%%%%%%%%%%%%%%%%%%%%%%

Symplectic scattering should be thought of as the classical counterpart of unitary scattering of quantum mechanics or quantum field theory. Classically, what is preserved by a scattering diffeomorphism is the basic structure of the theory, namely the symplectic structure of the manifold of classical states, whereas quantum mechanically, the scattering $S$-matrix has the property of preserving the fundamental structure of the space of quantum state vectors, namely the Hilbertian structure.

Given symplectic manifolds $(\cM_1,\omega_1)$ of ``in'' states, and $(\cM_2,\omega_2)$ of ``out'' states, we assume that a scattering process is given by a local symplectomorphism, viz., a local diffeomorphism 
\begin{equation}
S:\cM_1\to\cM_2
\qquad
\hbox{such that}
\qquad
\omega_1=S^*\omega_2.
\label{S}
\end{equation}
Such mappings being far from unique, we have to take into account the geometric features of the scattering device to try and find a unique symplectomorphism, $S$.

\goodbreak

In most cases, the ``in'' and ``out'' manifolds are Hamiltonian $G$-spaces\footnote{
These are symplectic manifolds $(\cM,\omega)$ equipped with a $G$-action $g\mapsto{}g_\cM$ for which $g_\cM^*\omega\equiv\omega$ and a  momentum mapping \cite{Sou}, i.e., a globally defined mapping $J:\cM\to\fg^*$ such that there holds $\omega(Z_\cM)=-d(J\cdot{}Z)$ for all $Z\in\fg$.
}
(e.g., coadjoint orbits) of some Lie group $G$, for instance a group of space(-time) automorphisms; they represent the free asymptotic states of the system. The scattering device reduces the original symmetry to a Lie subgroup $H\subset{}G$ whose action is assumed to intertwine the symplectomorphism $S$, that is
\begin{equation}
S\circ{}h_{\cM_1}=h_{\cM_2}\circ S
\label{H-equivariance}
\end{equation}
for all $h\in{}H$.
If $Z\in\fh$, where $\fh$ is the Lie algebra of $H$, we readily find from (\ref{H-equivariance}) that the associated fundamental vector fields are $S$-related, $Z_{\cM_2}=S_*Z_{\cM_1}$. This entails, via (\ref{S}) and (\ref{H-equivariance}), that $\omega_1(Z_{\cM_1})=S^*(\omega_2)(Z_{\cM_1})=S^*(\omega_2(S_*Z_{\cM_1}))=-S^*(d(J_2\cdot{}Z))$ where $J_2$ is the moment map of $(\cM_2,\omega_2,G)$. At last $\omega_1(Z_{\cM_1})=-d(S^*(J_2)\cdot{}Z)$ for all $Z\in\fh$ which, if $\cM_1$ and $\cM_2$ are connected, enables us to write the conservation law 
\begin{equation}
{J_1}{\vert\fh}=S^*(J_2{\vert\fh})
\label{J1=J2}
\end{equation}
that plays a central r\^ole in the determination of the sought scattering mapping, $S$.

%%%%%%%%%%%%%%%%%%%%%%%%%%%%%%%%%%%%%%%%%%%%%%%%%%%%%%%%%%%%%%%%%%%%%%%%%%%%%%%%%%%%
\subsection{Scattering of spinning light rays}
%%%%%%%%%%%%%%%%%%%%%%%%%%%%%%%%%%%%%%%%%%%%%%%%%%%%%%%%%%%%%%%%%%%%%%%%%%%%%%%%%%%%

Let us now turn to the effective computation of the scattering mapping of light rays with color $p$ and spin $s=\pm\hbar$ by the previously introduced interface separating two media of constant, unequal, refractive indices $n_1$ and $n_2$. 

From now on $(\cM_1,\omega_1)$ and $(\cM_2,\omega_2)$ will represent $\SE(3)$-coadjoint orbits respectively characterized by the invariants 
\begin{equation}
C_1=p_1^2,
\qquad
C'_1=p_1s_1,
\qquad
\&
\qquad
C_2=p_2^2,
\qquad
C'_2=p_2s_2,
\label{Casimir}
\end{equation}
where $p_1=p\,n_1$ $\&$ $p_2=p\,n_2$ with $p>0$ and $s_1,s_2\in\{+\hbar,-\hbar\}$. We handle, in this manner, all helicities at the same time. 

The canonical $2$-forms on $\cM_i\cong{}TS^2$ are given by (\ref{omegasp}) and read now
\begin{equation}
\omega_i(\delta\xi_i,\delta'\xi_i)=p_i\left[\la\delta\bu_i,\delta'\bq_i\ra-\la\delta'\bu_i,\delta\bq_i\ra\right]
-s_i\la\bu_i,\delta\bu_i\times\delta'\bu_i\ra,
\label{omega-i}
\end{equation}
for $i=1,2$.

Incoming light rays, i.e., hitting the interface in $M_1$, constitute a submanifold of $\cM_1$, whereas light rays refracted in $M_2$ form a submanifold of $\cM_2$. Moreover, reflection will be dealt with by considering $\cM_2=\cM_1$, as a manifold, whose symplectic $2$-form $\omega_2$ is defined by $p_2=p_1$ (since light bounces back in half-space $M_1$ with index $n_1$) and $s_2$.

As to the symmetry group of the optical interface, it is clearly given by the Lie subgroup 
\begin{equation}
H=\{(A,\bc)\in\SE(3)\,\vert\,A\bn=\bn,\la\bn,\bc\ra=0\},
\label{H}
\end{equation}
hence $H=\SE(2)\subset\SE(3)$.

We are now ready to implement (\ref{S}), (\ref{H-equivariance}) and (\ref{J1=J2}).

%-----------------------------------------------------------------------------------
\subsubsection{Conservation laws}
%-----------------------------------------------------------------------------------

The $H$-momentum mapping of $(TS^2,\omega)$ is the restriction ${J}{\vert\fh}$ of the Euclidean momentum mapping, $J$. We find \cite{Sou} that ${J(\bq,\bu)}{\vert\fh}=(L,\bP)$ is of the form 
\begin{equation}
\left\{
\begin{array}{rcl}
L&=&\la\bn,\bell\,\ra\\[6pt]
\bP&=&\bn\times\bp
\end{array}
\right.
\label{Jh}
\end{equation}
where $J(\bq,\bu)=(\bell,\bp)$ is as in (\ref{J}). 

If we put $(\bq_2,\bu_2)=S(\bq_1,\bu_1)$, the conservation law (\ref{J1=J2}) reads
\begin{eqnarray}
\label{L1=L2}
\la\bn,\bq_1\times\bp_1+s_1\bu_1\ra&=&\la\bn,\bq_2\times\bp_2+s_2\bu_2\ra\\
\label{P1=P2}
\bn\times\bp_1&=&\bn\times\bp_2
\end{eqnarray}
where 
\begin{equation}
\bp_i=p_i\bu_i
\label{bpi}
\end{equation}
for $i=1,2$.

Equation (\ref{P1=P2}) readily implies
\begin{equation}
\bp_2=\bp_1+\lambda\bn
\label{p2}
\end{equation}
where $\lambda$ is some a smooth function of $(\bq_1,\bu_1)$; taking into account the Euclidean invariants $C_1=\Vert\bp_1\Vert^2=p_1^2$ and $C_2=\Vert\bp_2\Vert^2=p_2^2$, see (\ref{Casimir}), already insures that $\lambda$ depends on
\begin{equation}
\alpha=\la\bn,\bp_1\ra
\label{alpha}
\end{equation}
only, via
\begin{equation}
\lambda^2+2\alpha\lambda+C_1-C_2=0.
\label{lambda2}
\end{equation}
Note that, with our orientation, incoming rays are such that $\alpha>0$.  We readily find the explicit expression\footnote{
If $C_1>C_2$, then $\alpha^2+C_2-C_1>0$ must furthermore hold true; if the latter condition is not satisfied, total reflection occurs.
}
\begin{equation}
\lambda=
\left\{
\begin{array}{ll}
-\alpha+\sqrt{\alpha^2+C_2-C_1}&\qquad(\hbox{refraction}, n_1\neq{}n_2)\\[10pt]
-2\alpha&\qquad(\hbox{reflection}, n_1=n_2)
\end{array}
\label{lambda}
\right.
\end{equation}
which will be used in the sequel.

\goodbreak

%-----------------------------------------------------------------------------------
\subsubsection{The scattering symplectomorphism}
%-----------------------------------------------------------------------------------

Taking advantage of (\ref{p2}), we now seek the diffeomorphism $S:(\bq_1,\bu_1)\mapsto(\bq_2,\bu_2)$ starting from the general Ansatz
\begin{equation}
\left\{
\begin{array}{rcl}
\bq_2&=&\bq_1+\mu\bp_1+\nu\bn+\varrho\bn\times\bp_1\\[6pt]
\bp_2&=&\bp_1+\lambda\bn
\end{array}
\right.
\label{Ansatz}
\end{equation}
where $\lambda$ is given by (\ref{lambda2}) and $\mu,\nu,\varrho$ are otherwise arbitrary functions of $(\bq_1,\bu_1)$.

$\bullet$ From (\ref{Ansatz}), (\ref{L1=L2}), (\ref{Casimir}), together with $C'/C=s/p$, we immediately obtain
$\alpha(C'_2/C_2-C'_1/C_1)-\varrho(C_1-\alpha^2)+\lambda{}C'_2/C_2=0$, or, if $\bn\times\bp_1\neq0$ (in the generic case of non normal incidence),
\begin{equation}
\varrho=\frac{1}{\Vert\bn\times\bp_1\Vert^2}\left[
\alpha\left(\frac{C'_2}{C_2}-\frac{C'_1}{C_1}\right)+\lambda\frac{C'_2}{C_2}
\right].
\label{rho}
\end{equation}
In the case of normal incident rays, $\Vert\bn\times\bp_1\Vert^2=C_1-\alpha^2=0$, we must have $(\lambda+\alpha)C'_2/C_2-\alpha{}C'_1/C_1=0$ with $\alpha=\sqrt{C_1}$ and $\lambda+\alpha=\sqrt{C_2}$ (resp. $\lambda+\alpha=-\sqrt{C_1}$) for refraction (resp. reflection). We therefore find
\begin{equation}
\left\{
\begin{array}{ll}
s_2=s_1&\qquad(\hbox{refraction})\\[10pt]
s_2=-s_1&\qquad(\hbox{reflection})
\end{array}
\label{polarization}
\right.
\end{equation}
which constitute non trivial conditions on the scattering symplectomorphism, $S$.

\goodbreak

$\bullet$ Take now into account the constraints 
\begin{equation}
\Vert\bp_i\Vert^2=C_i
\qquad
\&
\qquad
\la\bp_i,\bq_i\ra=0,
\label{constraints}
\end{equation}
for all $i=1,2$,
to further determine the yet unknown function $\nu$. If we put
\begin{equation}
z=\la\bn,\bq_1\ra
\label{z}
\end{equation}
for the $\bn$-component of ``position'' $\bq_1$, then $\la\bp_1,\bq_1\ra=\la\bp_2,\bq_2\ra=0$, together with~(\ref{Ansatz}) imply
\begin{equation}
\label{nu}
\nu=\frac{-1}{\alpha+\lambda}\left(
\lambda z+\mu(C_1+\alpha\lambda)
\right).
\end{equation}

$\bullet$ Let us use the previous Ansatz (\ref{Ansatz}) to express that $S:\xi_1\mapsto\xi_2$ is a symplectomorphism, namely $\omega_1(\delta\xi_1,\delta'\xi_1)=\omega_1(\delta\xi_2,\delta'\xi_2)$ for all $\delta\xi_i,\delta'\xi_i\in{}T_{\xi_i}\cM_i$, or
\begin{equation}
\begin{array}{c}
\la\delta\bp_1,\delta'\bq_1\ra-\la\delta'\bp_1,\delta\bq_1\ra
\displaystyle
-\frac{C'_1}{C_1^2}\la\bp_1,\delta\bp_1\times\delta'\bp_1\ra\\[10pt]
=\\[8pt]
\la\delta\bp_2,\delta'\bq_2\ra-\la\delta'\bp_2,\delta\bq_2\ra
\displaystyle
-\frac{C'_2}{C_2^2}\la\bp_2,\delta\bp_2\times\delta'\bp_2\ra
\end{array}
\label{EqSymplectomorphism}
\end{equation} 
for all tangent vectors compatible with the constraints (\ref{constraints}).

A tedious calculation shows us that (\ref{EqSymplectomorphism}) yields
\begin{eqnarray}
0
\nonumber
&=&
+\delta\lambda\delta'z-\delta'\lambda\delta{}z
+
\alpha(\delta\lambda\delta'\mu-\delta'\lambda\delta\mu)
+
\delta\alpha\delta'\nu-\delta'\alpha\delta\nu
+
\delta\lambda\delta'\nu-\delta'\lambda\delta\nu
\\%[8pt]
\label{tediousCalculation}
&&
+
\la\delta\bp_1,\bn\times\bp_1\ra\delta'(\varrho-C'_2/C_2^2\lambda)-\la\delta'\bp_1,\bn\times\bp_1\ra\delta(\varrho-C'_2/C_2^2\lambda)
\\%[8pt]
&&
\nonumber
-
\la(2\varrho+\lambda{}C'_2/C_2^2)\bn,\delta\bp_1\times\delta'\bp_1\ra
%\\%[8pt]
%&&
+
(C'_1/C_1^2-C'_2/C_2^2)\la\bp_1,\delta\bp_1\times\delta'\bp_1\ra.
\end{eqnarray}
In order to tackle (\ref{tediousCalculation}), we find it useful to introduce spherical coordinates $(\theta,\varphi)$ on the $2$-sphere described by $\bu_1=\bp_1/p_1=(\cos\varphi\sin\theta,\sin\varphi\sin\theta,\cos\theta)$.\footnote{
We have $\la\bu_1,\delta\bu_1\times\delta'\bu_1\ra=\sin\theta(\delta\theta\delta'\varphi-\delta'\theta\delta\varphi)$, and $\la\bn,\delta\bu_1\times\delta'\bu_1\ra=\cos\theta\sin\theta(\delta\theta\delta'\varphi-\delta'\theta\delta\varphi)$, and also $\la\bn,\bu_1\times{}d\bu_1\ra=\sin^2\theta\,d\varphi$.
}

\goodbreak

Rewrite
(\ref{tediousCalculation}) as
\begin{eqnarray}
0
\nonumber
&=&
+d\lambda\wedge{}dz
+
\alpha{}d\lambda\wedge{}d\mu
+
d\alpha\wedge{}d\nu
+
d\lambda\wedge{}d\nu
\\%[8pt]
\label{tediousCalculationBis}
&&
+
C_1\sin^2\theta(d\varphi\wedge{}d\varrho-C'_2/C_2^2\,d\lambda)
\\%[8pt]
&&
\nonumber
-
(2\varrho+\lambda{}C'_2/C_2^2)C_1\cos\theta\sin\theta\,d\theta\wedge{}d\varphi
%\\%[8pt]
%&&
+
(C'_1/C_1^2-C'_2/C_2^2)C_1^{3/2}\sin\theta\,d\theta\wedge{}d\varphi.
\end{eqnarray}

From (\ref{lambda2}) we get $d\lambda=-\lambda{}d\alpha/(\alpha+\lambda)$, while (\ref{alpha}) yields $d\alpha=-\sqrt{C_1}\sin\theta{}d\theta$. We then obtain the partial expression
$
d\lambda\wedge{}dz
+
\alpha{}d\lambda\wedge{}d\mu
+
d\alpha\wedge{}d\nu
+
d\lambda\wedge{}d\nu
=
-1/(\alpha+\lambda)d\alpha\wedge{}d(\lambda{}z+\alpha\lambda\mu-\alpha\nu)
=
-1/(\alpha+\lambda)^2d\alpha\wedge{}d((C_2-C_1)z+C_2\alpha\mu)
$
with the help of (\ref{nu}).
Some more effort is needed to finally transcribe (\ref{tediousCalculationBis}) as
\begin{equation}
\begin{array}{rcl}
0
&=d\alpha\wedge\Bigg[&
\displaystyle
d\left(\frac{(C_2-C_1)z+C_2\alpha\mu)}{(\alpha+\lambda)^2}\right)
-
\left(\frac{\lambda}{(\alpha+\lambda)}\frac{C'_2}{C_2}-\frac{C'_2}{C_2}+\frac{C'_1}{C_1}\right)d\varphi\\[14pt]
&&
+
\displaystyle
\left(
C_1\left(\frac{C'_2}{C_2^2}-\frac{C'_1}{C_1^2}\right)
-\alpha\lambda\frac{C'_2}{C_2^2}
+\lambda\frac{C'_2}{C_2^2}\frac{(C_1-\alpha^2)}{(\alpha+\lambda)}
\right)d\varphi
\Bigg].
\end{array}
\end{equation}
This readily implies
\begin{equation}
C_2-C_1+C_2\alpha\frac{\partial\mu}{\partial{}z}=0
\label{dmudz}
\end{equation}
and, with the help of (\ref{lambda2}), also gives
\begin{equation}
\frac{\partial\mu}{\partial\varphi}=0,
\label{dmudphi}
\end{equation}
which leaves us with
\begin{equation}
\mu=\frac{(C_1-C_2)}{C_2}\frac{z}{\alpha}+\widehat{\mu}=0
\label{mu}
\end{equation}
where $\widehat{\mu}=f(\alpha)$ is an arbitrary function of $\alpha$.

$\bullet$ So far, all four functions $\lambda,\mu,\nu,\varrho$ have been determined by (\ref{lambda}), (\ref{mu}), (\ref{nu}) and (\ref{rho}), up to an arbitrary function $\widehat{\mu}$. Let us show that, indeed, $\widehat{\mu}=0$.

Returning to the expression (\ref{Ansatz}) giving the scattering mapping which we write, for convenience,  $S:(\bq_1,\bp_1)\mapsto(\bq_2=\bq_1+\mu_1\bp_1+\nu_1\bn+\varrho_1\bn\times\bp_1,\bp_2=\bp_1+\lambda_1\bn)$, its inverse $S^{-1}:(\bq_2,\bp_2)\mapsto(\bq_1=\bq_2+\mu_2\bp_2+\nu_2\bn+\varrho_2\bn\times\bp_2,\bp_1=\bp_2+\lambda_2\bn)$ is such that $\lambda_2=-\lambda_1$ (where $\alpha_2=\alpha_1+\lambda_1$), $\mu_2=-\mu_1$, $\nu_2=\lambda_1\mu_1-\nu_1$ and $\varrho_2=-\varrho_1$. These relationships implement the principle of ray reversibility.

From the definition (\ref{z}), we get $z_2=z_1+\alpha_1\mu_1+\nu_1$ and find, resorting to~(\ref{mu}), that $\mu_1+\mu_2=\left[C_1\alpha_2^2f(\alpha_2)+C_2\alpha_1^2f(\alpha_1)\right]/C_1\alpha_2^2\equiv0$ iff $f=0$.

We obtain, at last,
\begin{equation}
\mu=\frac{(C_1-C_2)}{C_2}\frac{z}{\alpha}.
\label{muBis}
\end{equation}

We have thus completed the explicit determination of the scattering symplectomorphism by the plane interface separating two refracting media of constant indices.\footnote{The $H$-equivariance (\ref{H-equivariance}) of the unique symplectomorphism (\ref{Ansatz}), with (\ref{lambda}), (\ref{muBis}), (\ref{nu}) and (\ref{rho}),  can be directly checked to hold, the $H=\SE(2)$-action on $\cM=TS^2$ being given by $h_\cM(\bq,\bu)=(A\bq+\bc-A\bu\la{}A\bu,\bc\ra,A\bu)$, where $h=(A,\bc)\in{}H$ (see (\ref{H})).} 

\goodbreak

Let us collect and present the above findings in a new guise where the scattering mapping, $S$, is uniquely given by (\ref{Ansatz}) with (\ref{polarization}) and
\begin{equation}
\begin{array}{rcl}
\lambda
&=&
\left\{
\begin{array}{ll}
-\la\bn,\bp_1\ra+\sqrt{C_2-\Vert\bn\times\bp_1\Vert^2}&\qquad(\hbox{refraction})\\[10pt]
-2\la\bn,\bp_1\ra&\qquad(\hbox{reflection})
\end{array}
\right.\\[20pt]
\mu
&=&
\displaystyle
\left(\frac{C_1}{C_2}-1\right)\frac{\la\bn,\bq_1\ra}{\la\bn,\bp_1\ra}\\[14pt]
\nu
&=&
\displaystyle
\frac{C_1}{C_2}\lambda\frac{\la\bn,\bq_1\ra}{\la\bn,\bp_1\ra}\\[14pt]
\varrho
&=&
\displaystyle
\frac{1}{\Vert\bn\times\bp_1\Vert^2}\left[\left(\frac{C'_2}{C_2}-\frac{C'_1}{C_1}\right)\la\bn,\bp_1\ra+\frac{C'_2}{C_2}\lambda\right],
\end{array}
\label{coeffOK}
\end{equation}
the Casimir invariants being as in (\ref{Casimir}).

%%%%%%%%%%%%%%%%%%%%%%%%%%%%%%%%%%%%%%%%%%%%%%%%%%%%%%%%%%%%%%%%%%%%%%%%%%%%%%%%%%%%
\subsection{Snell-Descartes' laws of spinoptics \& Optical Hall Effect}
%%%%%%%%%%%%%%%%%%%%%%%%%%%%%%%%%%%%%%%%%%%%%%%%%%%%%%%%%%%%%%%%%%%%%%%%%%%%%%%%%%%%

Introducing the angle of incidence $\theta_1$ (resp. the scattering angle $\theta_2$) between $\bp_1$ (resp.~$\bp_2$) and $\bn$, we easily infer from~(\ref{P1=P2}) the law of refraction obeyed by the direction of light rays, namely $p_1\sin\theta_1=p_2\sin\theta_2$. As for the law of reflection, (\ref{lambda}) already yields the mirror transformation $\bp_1\mapsto\bp_2=\bp_1-2\bn\la\bn,\bp_1\ra$. 

\goodbreak

%-----------------------------------------------------------------------------------
\subsubsection{The generalized Snell-Descartes laws}\label{SDLawsSubsection}
%-----------------------------------------------------------------------------------

Summing up, and taking into account the specific result (\ref{polarization}), we write the Snell-Descartes laws of spinoptics as
\begin{equation}
\left\{
\begin{array}{lll}
n_2\sin\theta_2=n_1\sin\theta_1, 
&\qquad s_2=s_1,
&\qquad(\hbox{refraction})\\[10pt]
\theta_2=\pi-\theta_1,
&\qquad s_2=-s_1,
&\qquad(\hbox{reflection})
\end{array}
\right.
\label{SnellDescartes}
\end{equation}

Let us emphasize that these laws must be supplemented---as shown below---by a new law which unveils a phenomenon pertaining to geometrical spinoptics, namely a transverse shift of the scattered spinning light rays off the plane of incidence spanned by~$\bn$ and $\bp_1$ in generic position.

\goodbreak

Choose now as origin, $O$, of Euclidean space, the intersection of the in\-coming light ray and the interface separating the refractive media, so that $\bq_1=0$. From~(\ref{Ansatz}) and (\ref{coeffOK}), we get $\mu=0$ and $\nu=0$. 
We then obtain $\bq_2=\varrho\bn\times\bp_1$ where 
\begin{equation}
\varrho=
\displaystyle
\frac{1}{\Vert\bn\times\bp_1\Vert^2}\left[\frac{s_2}{p_2}\la\bn,\bp_2\ra-\frac{s_1}{p_1}\la\bn,\bp_1\ra\right].
\label{Shift}
\end{equation}
Note that there is no transverse shift in the case of normal incidence. We finally obtain the following expression for this transverse shift 
\begin{equation}
\bq_2-\bq_1=
\frac{\bn\times\bp_1}{\Vert\bn\times\bp_1\Vert}\frac{[s_2\cos\theta_2-s_1\cos\theta_1]}{p\,n_1\vert\sin\theta_1\vert}
\label{ShiftBis}
\end{equation}
which is clearly valid for either cases of refraction or reflection. This formula does formally agree with an analogous expression proposed by Onoda, Murakami and Nagaosa~\cite{OMN} who used quite a different viewpoint. 

Note, however, that the transverse shift (\ref{ShiftBis}) for \textit{reflected} rays vanishes in our framework; see also~\cite{Sou}. On the other hand, the nontrivial \textit{transverse shift} for \textit{refracted} spinning light rays, theoretically explained by (\ref{ShiftBis}) in the present context of geometrical spinoptics, is indeed a novel phenomenon known as the ``Optical Hall Effect'' and of a great importance in the new trends of experimental optics.

%-----------------------------------------------------------------------------------
\subsubsection{The special case of left-handed media}
%-----------------------------------------------------------------------------------

So far, we have been dealing with ordinary dielectric media. Quite interestingly, artificial materials enjoying a \textit{negative} refractive index (with both negative dielectric susceptibility and magnetic permeability) have been foreseen by Veselago a few decades ago \cite{Ves}. These brand new ``left-handed'' media (LHM) or metamaterials are nowadays manufactured in the laboratories and their strange optical properties systematically studied. See, e.g., \cite{BB04ter}.

Let us emphasize that our general formalism applies just as well in the presence of these LHM. For example, the Snell-Descartes laws, spelled out in Section~\ref{SDLawsSubsection}, still hold true if $n_2<0$, say. In this case, Equations (\ref{SnellDescartes}) account for ``negative reflection'', while (\ref{bpi}) shows that the linear momentum $\bp_2=p n_2\bu_2$ and the direction $\bu_2$ of a refracted ray are antiparallel.

One of the striking feature of these metamaterial is that the transverse shift of reflected and refracted spinning light rays may vanish identically, namely
\begin{equation}
\bq_1-\bq_2=0
\qquad
\mathrm{iff}
\qquad
n_1+n_2=0.
\label{ShiftBis}
\end{equation}
This characteristic property of metamaterials might provide a purely mechanical interpretation of the ``perfectness'' of LHM lenses that allow to break the Rayleigh limit of optical devices \cite{Pen}. 

%%%%%%%%%%%%%%%%%%%%%%%%%%%%%%%%%%%%%%%%%%%%%%%%%%%%%%%%%%%%%%%%%%%%%%%%%%%%%%%%%%%%
%%%%%%%%%%%%%%%%%%%%%%%%%%%%%%%%%%%%%%%%%%%%%%%%%%%%%%%%%%%%%%%%%%%%%%%%%%%%%%%%%%%%
\section{Conclusion and outlook}\label{Conclusion}
%%%%%%%%%%%%%%%%%%%%%%%%%%%%%%%%%%%%%%%%%%%%%%%%%%%%%%%%%%%%%%%%%%%%%%%%%%%%%%%%%%%%
%%%%%%%%%%%%%%%%%%%%%%%%%%%%%%%%%%%%%%%%%%%%%%%%%%%%%%%%%%%%%%%%%%%%%%%%%%%%%%%%%%%%

The basics of geometrical spinoptics have been laid to extend, from first principles, geometrical optics to spinning light rays. The point of view we have espoused made crucial use of  Euclidean geometry. By generalizing the Fermat prescription to the presymplectic manifolds upstream of generic coadjoint orbits of the Euclidean group, we have derived a $1$-dimensional foliation governing the trajectories of spinning light rays in arbitrary dielectric media. A refinement of the classic Snell-Descartes laws readily followed, together with the expression of the local scattering symplectomorphism undergone by spinning refracted and reflected light rays. This enabled us to derive a formula for the associated transverse shift, specific to the Optical Hall effect. 

A number of queries, triggered by the present study, remain the subject of future work; let us mention here some few examples.

Revisiting the theory of caustics within this new framework would certainly be a worthwhile task, in view of the refinement the Snell-Descartes laws governed by noncommutativity of the wave plane.

Quantizing geometrical spinoptics is also serious endeavor. One might profit by the fact that prequantization of the symplectic manifold of photonic states, with $s=\pm\hbar$, is given by the contact structure on the quotient $\SO(M)/(\ker(\varpi)\cap\ker(d\varpi))$ defined by (\ref{varpispMg}). Another route to quantization might, alternatively, be offered by the procedure of conformally equivariant quantization~\cite{DO}. It would also be interesting to see how close to Maxwell theory should such a quantization lead us.

Another challenging project would be to tackle all polarization states at a single stroke by considering that circular polarization states given by the coadjoint orbits of the Euclidean group are, in fact, the building blocks or elementary systems of a more elaborate, quantum, theory of spinoptics.

\goodbreak

At last, it seems reasonable to envisage extending geometric spinoptics to the case of Faraday-active media by coupling, from the outset, the photon spin and the external magnetic field via the color, $p$, much in the same way as the (anomalous) magnetic interaction term is introduced via the mass in a general relativistic framework \cite{Duv,Sou2,Ste,DH}.

%\newpage

%%%%%%%%%%%%%%%%%%%%%%%%%%%%%%%%%%%%%%%%%%%%%%%%%%%%%%%%%%%%%%%%%%%%%%%%%%%%%%%%%%%%
%%%%%%%%%%%%%%%%%%%%%%%%%%%%%%%%%%%%%%%%%%%%%%%%%%%%%%%%%%%%%%%%%%%%%%%%%%%%%%%%%%%%


\begin{thebibliography}{99}
%%%%%%%%%%%%%%%%%%%%%%%%%%%%%%%%%%%%%%%%%%%%%%%%%%%%%%%%%%%%%%%%%%%%%%%%%%%%%%%%%%%%
%%%%%%%%%%%%%%%%%%%%%%%%%%%%%%%%%%%%%%%%%%%%%%%%%%%%%%%%%%%%%%%%%%%%%%%%%%%%%%%%%%%%

\bibitem{Arn}
V.~I.~Arnold, 
\textsl{Mathematical Methods of Classical Mechanics},
Second Edition, Springer (1989).

\bibitem{BM}
A. B\'erard and H. Mohrbach,
"NonAbelian Berry Phase in Noncommutative Quantum Mechanics",
\texttt{arXiv:hep-th/0404165}.


\bibitem{Bli}
K. Yu. Bliokh,
``Geometrical optics of beams with vortices:
Berry phase and orbital angular momentum Hall effect'',
\texttt{arXiv:physics/0603072}.

\bibitem{BB04}
K. Yu. Bliokh and Yu. P. Bliokh, 
``Topological spin transport of photons: the optical Magnus Effect and Berry Phase'', 
Phys. Lett. A {\bf 333}, 181--186, (2004).

\bibitem{BB04bis}
K. Yu. Bliokh and Yu. P. Bliokh, 
``Modified geometrical optics of a smoothly inhomogeneous isotropic medium: the anisotropy, Berry phase, and the optical Magnus effect'', 
Phys. Rev. E {\bf 70}, 026605, (2004).
%\texttt{arXiv:physics/0402014}.

\bibitem{BB04ter}
K. Yu. Bliokh and Yu. P. Bliokh,
``What are the left-handed media and what is interesting about them?'',
Physics-Uspekhi {\bf 47}, 393--400, (2004).

\bibitem{BB05}
K. Yu. Bliokh and Yu. P. Bliokh, 
``Conservation of Angular Momentum, Transverse Shift, and Spin Hall Effect in Reflection and Refraction of Electromagnetic Wave Packet'', 
Phys. Rev. Lett. {\bf 96}, 073903, (2006)
%\texttt{arXiv:physics/0508093}.

\bibitem{BF}
K. Yu. Bliokh and V. D. Freilikher, 
``Topological spin transport of photons: Magnetic monopole gauge field in Maxwell's equations and polarization splitting of rays in periodically inhomogeneous media'', 
Phys. Rev. B {\bf 72}, 035108, (2005).

%\bibitem{BW}
%M. Born and E. Wolf,
%\textsl{Principles of optics},
%Cambridge University Press (1999). 

\bibitem{Bou}
D. G. Boulware, 
``Phase-Shift Analysis of the Translation of Totally Reflected Beams'', 
Phys. Rev. D {\bf 7}, 2375--2382, (1973).

\bibitem{CN}
J. F. Cari$\tilde\mathrm{n}$ena and N. Nasarre, 
``Presymplectic geometry and Fermat's principle for anisotropic media'', 
J. Phys. A {\bf 29}, 1695--1702, (1996).
%\texttt{arXiv:hep-th/9602037}.

\bibitem{CN2}
J. F. Cari$\tilde\mathrm{n}$ena and N. Nasarre, 
``On the symplectic structure arising in geometric optics'', 
Fortschr. Phys. {\bf 44}, 181--198, (1996).

\bibitem{Cos}
O. Costa de Beauregard, 
``Translational Inertial Spin Effect with Photons'', 
Phys. Rev. {\bf 139}, 1443--1446, (1965).

\bibitem{Dix}
W. G. Dixon,
``On a classical theory of charged particles with spin and the classical limit of the Dirac equation'', Il Nuovo Cimento {\bf 38}, 1616--1643, (1965).

\bibitem{Duv}
C. Duval,
``The general relativistic Dirac-Pauli particle: an underlying classical model'',
Ann. Inst. Henri Poincar\'e {\bf 25 A}, 345--362, (1976).

\bibitem{DE}
C. Duval and J. Elhadad,
``Geometric Quantization and Localization of Relativistic Spin Systems'',
in \textsl{Proc. AMS of the 1991 Joint Summer Research
Conference on Mathematical Aspects of Classical Field Theory},
Seattle 1991,
(M.J.~Gotay, J.E.~Marsden, V.E.~Moncrief Eds),
Contemporary  Mathematics {\bf 132}, 317--330, (1992).


\bibitem{DET}
C. Duval, J. Elhadad, and G.M.~Tuynman,
``Pukanszky's Condition and Symplectic Induction'',
J. Diff. Geom.
{\bf 36}, 331--348, (1992).

\bibitem{DFS}
C. Duval, H. H. Fliche, and J.-M. Souriau,
``Un mod\`ele de particule \`a spin dans le champ gravitationnel et \'electromagn\'etique'',
Comptes Rendus de l'Acad\'emie des Sciences de Paris,
S\'erie A. {\bf 274}, 1082--1084, (1972).

\bibitem{DH}
C.~Duval and P.~Horv\'athy,
``Anyons with Anomalous Gyromagnetic Ratio and the Hall effect'',
Phys. Letters B {\bf 594}, 402--409, (2004).

\bibitem{DHH}
C.~Duval, Z.~Horv\'ath, and P.~Horv\'athy,
%``Fermat Principle for spinning light and the Optical Hall effect'',
``Fermat Principle for spinning light'',
\texttt{arXiv:cond-mat/0509636}.

\bibitem{DO}
C. Duval and V. Ovsienko,
``Conformally Equivariant Quantum Hamiltonians'',
Selecta Math. New Ser. {\bf 7}, 291--320, (2001).

\bibitem{GBM}
P. Gosselin, A. B\'erard, and H. Mohrbach,
``Spin Hall effect of photons in a static gravitational field'',
\texttt{arXiv:hep-th/0603227}

\bibitem{GS}
V. Guillemin and S. Sternberg,
\textsl{Symplectic techniques in physics},
Cambridge University Press, Cambridge (1984).

\bibitem{GS2}
V. Guillemin and S. Sternberg,
``Souriau scattering and the Yang-Mills dust'',
Ann. Physics {\bf 165}, 259--279, (1985).

\bibitem{Igl}
P. Iglesias, 
\textsl{Sym\'etries et moment}, Hermann (2000). 

\bibitem{Imb}
C. Imbert, 
``Calculation an Experimental Proof of the Transverse Shift Induced by Total Internal Reflection of Circularly Polarized Light Beam'', 
Phys. Rev. D {\bf 5}, 787--796, (1972).

\bibitem{Kir}
A. A. Kirillov, 
%\textsl{El\'ements de la th\'eorie des repr\'esentations}, 
%Editions Mir, Moscou, 1974. 
\textsl{Elements of the theory of representations},
Springer-Verlag, Berlin-New York, (1976). 

\bibitem{Kos}
B. Kostant, 
``Quantization and Unitary representations, Part I'', 
Lecture Notes in Math \textbf{170}, p. 87--208, Springer-Verlag (1970).

\bibitem{Kun}
H. P. K\"unzle,
``Canonical dynamics of spinning particles in gravitational and electromagnetic fields'',
J. Math. Phys. {\bf 13}, 739--744, (1972).

\bibitem{LZ}
V. S. Liberman and B. Ya. Zel'dovich,
``Spin-orbit interaction of a photon in an inhomogeneous medium'', 
Phys. Rev. A {\bf 46}, 5199--5207, (1992).

\bibitem{MR}
J.-E. Marsden and T. Ratiu, 
\textsl{Introduction to Mechanics and Symmetry}, 
Springer (1999).

\bibitem{OMN}
M. Onoda, S. Murakami, and N. Nagaosa, ``Hall Effect of Light'', Phys. Rev. Lett. {\bf 93}, 083901, (2004).

\bibitem{Pap}
A. Papapetrou, 
``Spinning test particles in General Relativity.1.'',
Proc. Roy. Soc. {\bf A 209}, 248--258, (1951).

\bibitem{Pen}
J. B. Pendry,
``Negative Refraction Makes a perfect Lens'',
Phys. Rev. Lett. {\bf 85}, 3966--3969, (2000).

\bibitem{Sou}
J.-M.~Souriau,
\textsl{Structure des syst\`emes dynamiques}, Dunod (1970, \copyright 1969);
\textsl{Structure of Dynamical Systems. A Symplectic View of Physics},
translated by C.H.~Cushman-de Vries (R.H.~Cushman and G.M.~Tuynman,
Translation Editors), 
Birkh\"auser (1997).

\bibitem{Sou2}
J.-M.~Souriau, ``Mod\`ele de particule \`a spin dans le champ 
\'electromagn\'etique et gravitationnel'',
Ann. Inst. Henri Poincar\'e, {\bf 20 A}, 315--364, (1974).

\bibitem{Ste}
S. Sternberg, 
``On the Role of Field Theories in our Physical Conception of Geometry'',
in Proc. 2nd Bonn Conf.
\textsl{Diff. Geom. Meths. in Math. Phys},
Lecture Notes in Mathematics, {\bf 676}, p. 1--80, Springer-Verlag (1978).

\bibitem{Tig}
B. A. van~Tiggelen,
``Transverse diffusion of light in Faraday-Active Media'', 
Phys. Rev. Lett. {\bf 75}, 422--424, (1995).

\bibitem{Ves}
V. G. Veselago, 
``The electrodynamics of substances with simultaneously negative values of $\epsilon$ and $\mu$'',
Sov. Phys. Usp. {\bf 10}, 509--514, (1968).

\bibitem{WSRLvT}
S. Wiebel, A. Sparenberg, G. L. J. A. Rikken, D. Lacoste, and B. A. van~Tiggelen,
``Photonic Hall effect in absorbing media'', 
Phys. Rev. E {\bf 62}, 8636--8639, (2000).





\end{thebibliography}
\end{document}